\documentclass{tMOP2e}

\usepackage{amsmath}
\usepackage{hyperref}
\usepackage{amssymb}
\usepackage{color}
\usepackage{graphicx}
\usepackage{latexsym}
\usepackage{multirow}
\usepackage{xspace}
\usepackage{textcomp}
\usepackage{subfigure}

\newcommand{\gws}{gravitational waves\xspace}
\newcommand{\gw}{gravitational wave\xspace}
\newcommand{\ee}[1]{\!\times\!10^{#1}}

\title{Advanced technologies for future ground-based,
 laser-interferometric gravitational wave detectors}

\author{Giles Hammond$^{a}$ $^{\ast}$\thanks{$^{\ast}$ 
Email: giles.hammond@glasgow.ac.uk}, 
Stefan Hild$^{a}$ $^{\dagger}$\thanks{$^{\dagger}$ 
Email: stefan.hild@glasgow.ac.uk} and 
Matthew Pitkin$^{a}$ $^{\ddagger}$\thanks{$^{\ddagger}$ 
Email: matthew.pitkin@glasgow.ac.uk \vspace{6pt}} \\\vspace{6pt} 
$^{a}${\em{SUPA, School of Physics and Astronomy, 
University of Glasgow, University Avenue, Glasgow, G12 8QQ, UK}} }

\begin{document}

\maketitle

\begin{abstract}
We present a review of modern optical techniques being used and developed for the field of gravitational wave
detection. We describe the current state-of-the-art of gravitational waves detector technologies with regard to
optical layouts, suspensions and test masses. We discuss the dominant sources and noise in each of these subsystems
and the developments that will help mitigate them for future generations of detectors. We very briefly summarise some of
the novel astrophysics that will be possible with these upgraded detectors.
\end{abstract}

\section{Gravitational waves: How to observe the most violent events in the Universe?}\label{sec:gws}
Over the last century our ability to view the Universe has undergone many revolutions. Observational astronomy has
progressed from a time when the only tool was ground-based visible light photometry to an era of ground-and-space-based
multi-wavelength photometry and spectroscopy. These advances have led to countless new discoveries and greatly increased
our knowledge of the Universe, but have only been possible through major technological and scientific leaps in
instrumentation. Sometimes this has been through serendipitous applications from different fields, but often it has been
the astronomical need that has pushed the technological boundaries.

\begin{figure}
\begin{center}
\includegraphics[width=1.0\textwidth]{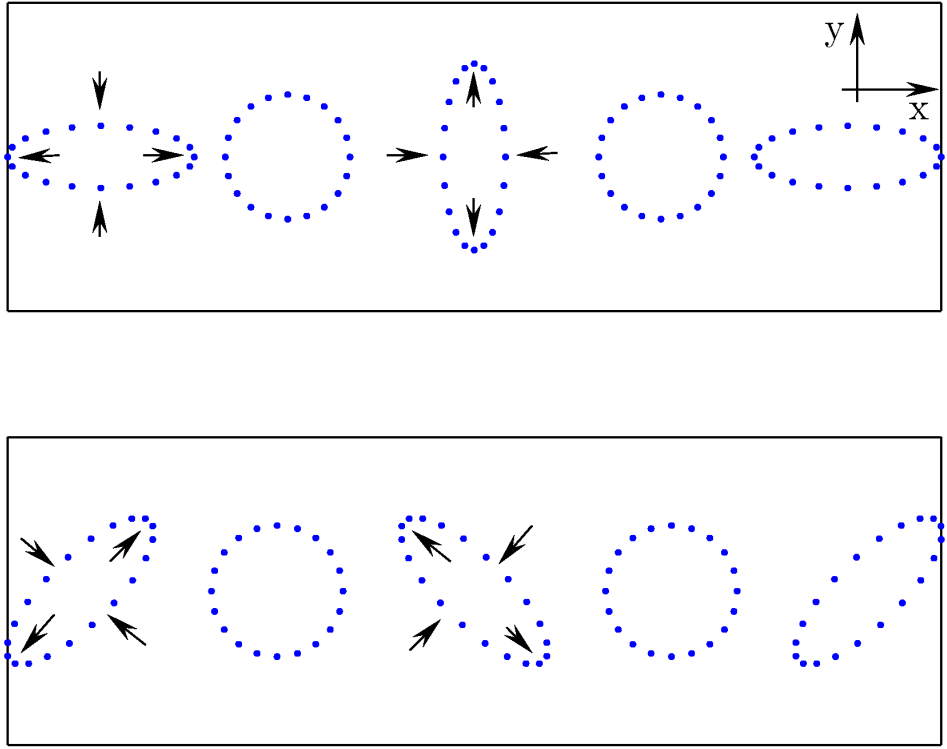}
\caption{The effect of a passing gravitational wave on a ring of freely falling test particles. The top and bottom
panels show the effect of the `plus' and `cross' polarisation modes respectively.}
\label{fig:ringofparticles}
\end{center}
\end{figure}

Gravitational waves (also called gravitational radiation) offer another new window on the Universe moving beyond the
electromagnetic spectrum. They are a direct prediction of Einstein's General Theory of Relativity (GR), although their
main properties can be inferred using Newtonian gravity and the principal of special relativity that information
(including the influence of gravity) cannot propagate faster than the speed of light \cite{Schutz:1984}. As they travel
through space gravitational waves manifest as a transverse varying tidal strain, which would have the effect of altering
the proper distance between freely falling objects. Gravitational waves are quadrupolar in nature and have two
polarisation states (called `plus' [$+$] and `cross' [$\times$]) rotated by 45$^\circ$ with respect to each other (see
Figure~\ref{fig:ringofparticles}). These states have orthogonal polarisation.

Gravity is the weakest of the fundamental forces and correspondingly the effect of \gws is tiny. Their strength is often
measured as a strain in space, where for our ring of particles in Figure~\ref{fig:ringofparticles} a maximum
displacement of $\Delta l$, given an un-displaced radius of $l$, means a strain of $h = \Delta l/l$. For a source at a
distance $r$ the approximate strain amplitude is given by
\begin{equation}
h \approx \frac{2GM}{c^2r}\left(\frac{v}{c}\right)^2,
\end{equation}
where $v$ is the non-spherical component of the source's motion (spherically symmetric motions, e.g.\ a rotating perfect
sphere will not radiate as they lack a time-varying quadrupole moment) and $M$ is the mass involved in that motion. It
can be seen that even for nearby sources (e.g.\ $1\,{\rm kiloparsec} = 3.09\ee{19}$\,m) with the mass of the Sun
(1$M_{\odot} = 1.99\ee{30}$\,kg) moving at near-relativistic velocities ($0.01c = 3.00\ee{6}$\,m/s) the strain at Earth
will be $h \sim 3\ee{-20}$, equivalent to a displacement between the Sun and the Earth of $\sim 5$\,nm.

As of yet no \gws have been directly detected. However, their existence has long been indirectly inferred from the
excellent agreement between theory and observation of the orbital decay of neutron star binary systems (e.g.\
\cite{Taylor:1982}).

\subsection{Astrophysical sources of gravitational waves}
Any non-axisymmetric acceleration of mass will emit \gws, but as seen above the emission of strong (and potentially
observable) \gws requires very large masses and relativistic motion. To fulfil both these requirements the sources must
be compact and dense. The archetypal source is the coalescence of a pair of compact stars (e.g.\ neutron stars or black
holes) in a tight binary system. Other sources include: individual rotating neutron stars with sustained distortions on
them \cite{Zimmermann:1979,Bonazzola:1996}; vibrating neutron stars \cite{Andersson:1998} and black holes
\cite{Teukolsky:1973}; the collapsing core of a star during a supernova \cite{Ott:2009} or $\gamma$-ray burst
\cite{Corsi:2012}; or, more exotically, inflation following the Big Bang \cite{Rubakov:1982} and cosmic strings
\cite{Vilenkin:1981,Damour:2000}. An excellent review of \gw sources and the science that can be learned from them is
presented in \cite{Sathyaprakash:2009}.

\subsection{The detection of gravitational waves}
The basic principle behind any \gw detector is to measure the relative displacement of freely falling bodies. The first
attempts to construct a detector to observe them were made by Joseph Weber in the 1960s \cite{Weber:1969}. He used
resonant mass, or ``bar'', detectors consisting of a suspended aluminium cylinder encircled with piezoelectric sensors,
and designed to resonate within a narrow frequency band when excited by a passing \gw. Modern bar detectors utilised
improved seismic isolation and operated at cryogenic temperatures ($\leq6$\,K) to reduce the thermal noise. Detectors
including ALLEGRO (Louisiana, US), EXPLORER (CERN), NAUTILUS (Frascati, Italy) and NIOBE (Perth, Australia) have been
used to set interesting limits on gravitational wave emission \cite{Ronga:2006}. Following on from these inherently
narrow-band ``bar'' detectors came the broad-band interferometric \gw detectors \cite{Moss:1971, Weiss:1972,
Drever:1977, Billing:1979}, which were pioneered by Rai Weiss at MIT \cite{Weiss:1972}. The aim of these was to use
interferometry to measure the displacement of two suspended mirror-coated test masses at the end of each arm. The
techniques required to allow these detectors to achieve the extraordinary accuracy necessary to detect a \gw will be
detailed within this review. One thing that can be seen straight away from the previous definition of the strain, $h$,
is that larger displacements, or smaller strains, can be measured if the separation of the test masses is large.
Therefore a prerequisite of current and future generations of interferometric \gw detectors is that the arm lengths are
long, generally on the few kilometres scale.

\subsection{Searches for gravitational waves}\label{sec:gwsearches}
The most recent efforts to detect \gws have come from the US Laser Interferometric Gravitational-wave Observatory (LIGO)
\cite{Abramovici:1992, Abbott:2009b}, the French-Italian Virgo \cite{Bradaschia:1990, Acernese:2006} and the
British-German GEO600 \cite{Willke:2007} projects. LIGO operates two sites (one in Livingston, Louisiana called LLO and
one in Hanford, Washington called LHO) each housing an interferometer with 4\,km arms and Fabry-Perot cavities
(previously LHO also housed a 2\,km interferometer). Virgo is a 3\,km interferometer with Fabry-Perot cavities in
Cascina, Italy. GEO600 (or latterly GEO-HF \cite{Willke:2006, Lueck_HF}), located near Hannover, Germany, has 1.2\,km 
arms, but in a folded configuration and with no Fabry-Perot cavities.

Over the last decade these detectors have produced unprecedented limits on \gw emission from a wide variety of sources.
The sensitivity of these detectors has meant that they would be able to observe neutron star binary coalescences out to
distances of a few tens of Mpc \cite{Abadie:2010} (a review of expected source rates can be found in
\cite{Abadie:2010b}). Some highlights of the astrophysics achieved include: surpassing the spin-down limit (assuming the
star's observed slow down is entirely due to loss of energy through \gws) for the Crab \cite{Abbott:2008, Abbott:2010}
and Vela \cite{Abadie:2011} pulsars; excluding binary coalescences as progenitors of two nearby $\gamma$-ray bursts
\cite{Abbott:2008b, Abadie:2012}; and, limiting the stochastic \gw background to have an energy density less than
$7\ee{-6}$ of the closure density of the Universe \cite{Abbott:2009}. A review of many more of the searches can be found
in \cite{Pitkin:2011}.

The LIGO and Virgo detectors are currently undergoing upgrades to their ``advanced'', or second generation,
configurations: Advanced LIGO (aLIGO) \cite{Harry:2010} and Advanced Virgo (AdV) \cite{Acernese:2009}. An estimate of
the commissioning and observation schedule for these advanced detectors is discussed in \cite{Aasi:2013}. Another second
generation interferometric detector, which will be located underground and utilise cryogenic cooling of its mirror, is
the Japanese KAGRA \cite{Somiya:2011} (formerly the Large-Scale Cryogenic Gravitational-wave Telescope, or LCGT
\cite{Kuroda:2010}). These detectors will aim to increase sensitivity over the initial generation of detectors by a
factor of ten, giving a thousandfold increase the observable volume, in addition to lowering the operating frequency
down to $\simeq 10$\,Hz. Theoretical predictions suggest \cite{Abadie:2010b} that the likely number of observable
neutron stars coalescences for these detector, once at design sensitivity, could be 40 per year, with the most
optimistic estimates suggesting over one per day.

One fundamental limit to the low-frequency ($\lesssim 1$\,Hz) sensitivity of ground-based detectors is Newtonian noise,
so to observe sources below this (e.g.\ supermassive black hole binaries \cite{Haehnelt:1994} and extreme mass ratio
black hole inspirals \cite{Gair:2004}) requires detectors in space like the proposed evolved
Laser Interferometer Space Antenna
(eLISA) \cite{Seoane:2013, Jennrich:2009}, or similar (e.g.\ \cite{Kawamura:2011, Harry:2006, Ni:2009}). Detection of
nanoHz \gws from the most massive supermassive black hole systems is also possible via use of high-precision pulsar
timing \cite{Hobbs:2010}, in which pulsars themselves are used as the test masses for a
galactic-scale detector. Even
lower frequency \gws could be imprinted in the Cosmic Microwave Background (CMB) and be detectable through the
polarisation state of the microwave photons. Recent observations by the BICEP2 Collaboration show the first possible
evidence for such relic gravitational waves in the CMB \cite{Ade:2014}, although confirmation with other independent
observations would strengthen the detection (e.g.\ with the Planck satellite \cite{Ade:2013, Zhao:2010}).
We will not discuss any of these techniques any further in this review.

\section{State-of-the-art Interferometer for Gravitational Wave detection}\label{sec:aLIGO_conf}
\begin{figure}
\begin{center}
\includegraphics[width=0.6\textwidth]{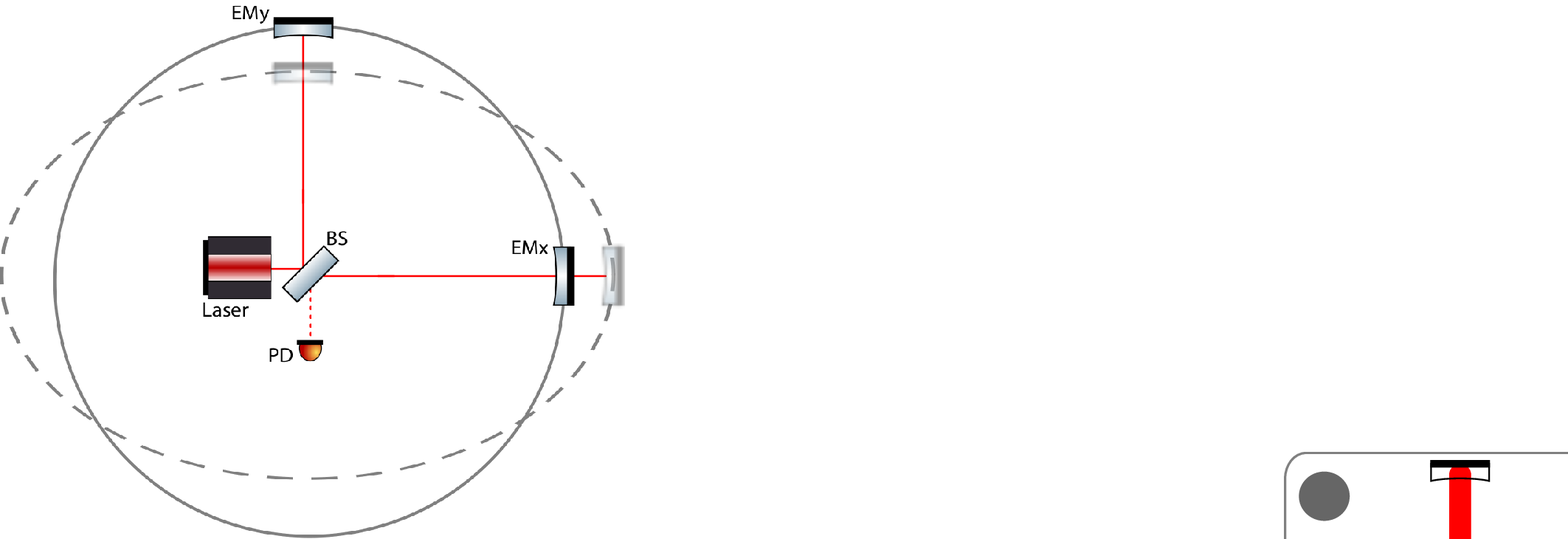}
\caption{Simplified schematic of Michelson interferometer acting as a gravitational wave detector and 
the effect of a $+$ polarised wave passing perpendicular through the plane of the interferometer. The light
from a laser is split 50:50 by a beam splitter (BS) and enters the two  arms of the interferometer. 
After bouncing back from the end mirrors (EMx and EMy) the light finally recombines on the BS. The 
 interference pattern is detected by a photodiode (PD). In the presence of a  
gravitational wave, one arm is increased in length the other shortened, resulting for a
suitably strong gravitational wave in a detectable change of the interference 
pattern.}
\label{fig:ifo_GW}
\end{center}
\end{figure}

In this section we will give a brief overview of the interferometer layout and configuration of a state-of-the art
gravitational  wave detector, such as aLIGO \cite{Harry:2010}.  Fundamentally all these instruments are based on large 
Michelson interferometers  continuously measuring the differential fluctuations in space-time of the two perpendicular
interferometer arms,  by actually monitoring the light travel time between high-quality mirrors acting as test masses. 
Due to the fact that gravitational radiation is of quadrupolar nature (see Figure~\ref{fig:ringofparticles}) a
Michelson interferometer is well suited for observing them.  Consider a simple Michelson interferometer as displayed in
Figure~\ref{fig:ifo_GW}. If, for instance, a gravitational wave of $+$-polarisation passing perpendicular to the
interferometer plane through the Michelson interferometer, causes one of the interferometer arms to become shortened,
while at the same time the other one will be increased in length. The change in the relative travel time through both
arms leads, when the returning beams recombine at the beam splitter,  to a change in the interference pattern which can
be detected by a photodiode at the output port of interferometer. Carrying out such a differential measurement of the
length of the two arms is beneficial, because it allows one to suppress a variety of technical noise sources common in
both arms, such as frequency noise of the laser. 
 
 State-of-the-art gravitational wave observatories are extremely complex machines, consisting of kilometer 
 long ultra-high vacuum systems, ultra-stable high-power continuous wave lasers, high-performance seismic isolation
systems and dozens of super-polished low-loss mirrors suspended in multi-cascaded
 pendulum systems (see for instance \cite{Harry:2010, fritschel:2007}). To establish low-noise
operation these instruments require 
 hundreds of control loops to be closed, making, for example, sure that all 
 mirrors are kept on their longitudinal operation point and are aligned exactly in 
 the right direction. 
 
Though these gravitational wave detectors are technically very complex, their fundamental working principles
can be broken down into two general requirements. To build a sensitive
 gravitational wave detector first of all one needs to make  sure that the test
  masses are more quiet than the expected effect on the arm length from a passing
  gravitational  wave. This is the reason why we host our instruments inside
 vacuum systems and take great care to reduce the  thermal noise of the mirrors
and their suspensions to an equivalent displacement noise below $10^{-20}$m/$\sqrt{\rm Hz}$ at a
frequency of 100\,Hz. Secondly, one needs to ensure to be able to read out
the mirror positions, i.e.\ the differential arm length degree of freedom to 
the required accuracy, while at the same time avoiding introducing any significant 
level of  back-action noise from the measurement process.  As we will see below 
this point directly relates to the interferometric readout of the test mass 
position and the associated so-called \emph{quantum noise}
 (see section \ref{sec:QN}). In addition to these 
fundamental sources there are a variety of technical noise sources, which are kept 
$\lesssim 10$\,\% of the fundamental noise sources, including laser frequency and 
intensity noise, charging noise, scattered light and the facility limits due to 
residual gas pressure in the vacuum system \cite{Abbott:2009b}. The expected limit to
performance from the basic noise sources of the aLIGO detector is shown in Figure~\ref{strain:aLIGO}. In order 
to meet the noise requirement of the advanced detectors a variety of techniques 
are used to mitigate the fundamental noise sources as described in 
Section~\ref{sec:noise}.

\subsection{Laser sources for \gw detectors}

In order to achieve their targeted sensitivity \gw detectors employ ultra-stable, 
high-power, single-mode, continuous wave laser sources. Good stability is 
required in terms of low beam jitter noise, high frequency stability and low 
relative intensity noise. It is worth noting that for some of these noise sources,
a good stability is not only required in the frequency range of the \gw detection
band, but also at DC as well as at radio frequencies (RF). For instance the relative 
intensity noise has to be below a certain level ($<10^{-8}/\sqrt{\rm Hz}$) in the audioband, because otherwise
the power noise would couple either directly to the \gw channel or via raidtion
pressure acting on the main test masses. At the same time it is important that
the DC power level is very stable over minutes to months time scales to ensure
stability of all control loops keeping the mirrors at their operation points.
Furthermore, it is also important to have low noise ($<10^{-9}/\sqrt{\rm Hz}$) in the RF range
(about 10\,MHz to 100\,MHz), as otherwise the noise could couple via the 
RF modulation sidebands  (see Sections~\ref{sec:length} and \ref{sec:readout}) 
to the \gw channel. Another requirement for laser sources to be suitable for 
\gw detectors is the provision of linear, fast and wide-range actuators allowing external frequency
and power stabilisation.
 
While early \gw detector prototype interferometers used Argon ion lasers, 
nowadays all \gw interferometers employ solid state lasers emitting 
at a wavelength of 1064\,nm. The preferred transverse electromagnetic mode is 
the TEM$_{00}$ mode, i.e.\ a Gaussian beam. Advanced gravitational wave detectors require
laser output powers of the order of 100\,W to 200\,W. 

As a good example of a state-of-the-art laser source, we will in the following briefly 
describe the Advanced LIGO laser systems \cite{Winkelmann,Kwee:12}. This  
3-stage system can deliver about 150\,W with more than 99.5\,\% of the power being
in the TEM$_{00}$ mode. The first stage consists of a non-planar ring-oscillator 
(NPRO) based on a Nd:YAG laser crystal, pumped at a wavelength of 808\,nm. The 
NPRO provides an output power of only 2\,W, but due to its monolithic resonator 
design, it has very low intrinsic frequency noise. The following two stages of 
the laser system are used to increase the output power, while at the same time 
inheriting the frequency stability and low noise of the NPRO. The second stage
consists of a Nd:YVO$_4$ amplifier with an output power of 35\,W. Finally, at the
end of the chain a high-power,  injection locked, ring-oscillator with four Nd:YAG 
crystals provides a maximum output power of more than 200\,W.

\subsection{Mode cleaning} \label{sec:MC}

\begin{figure}
\begin{center}
\includegraphics[width=0.6\textwidth]{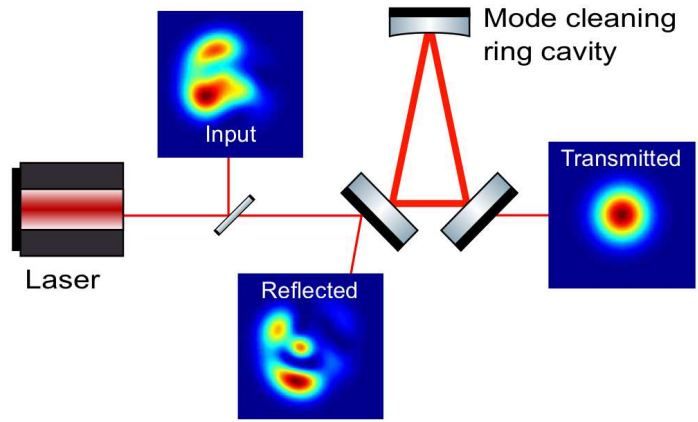}
\caption{Sketch illustrating the working principle of mode cleaning cavity.
 The ring resonator is designed to be resonant for the TEM$_{00}$ mode, but reject higher 
 order TEM mode light.}
\label{fig:MC}
\end{center}
\end{figure}

In order to provide the required spatial and temporal stability of the laser beam 
entering the main interferometer all 
currently operating \gw detectors make use of mode cleaning cavities \cite{MC:Albrecht,Willke:98}.
A high-finesse, 3-mirror ring-cavity is used to filter out geometric
noise in the laser beam, i.e. to suppress beam jitter and higher 
order optical modes.

Figure \ref{fig:MC} illustrates the working principle of a mode cleaner 
cavity. The light incident onto the ring cavity (Input) is not a pure 
TEM$_{00}$ beam, but can be strongly distorted, i.e. contain significant
amounts of higher order modes, especially first and second order modes.
The length of the mode cleaner cavity is set so that the fundamental mode
of the beam is resonant. Since the Guoy phase shift accumulated in a round trip
through the ring cavity depends on the optical mode order, the geometry of the 
mode cleaning cavity can be designed to be resonant only for the zero-order beam
and off resonance for higher order beams. As a result the input light is split
off in two components: While the TEM$_{00}$ mode is resonating in the mode cleaner
and being transmitted towards the main interferometer, the higher order modes
are reflected from the cavity. The same effect could also be achieved by 
a linear 2-mirror cavity. However using a 3-mirror cavity allows an easy 
separation of the input beam and the reflected `junk' light.
 
For a triangular mode cleaner cavity as displayed in Figure~\ref{fig:MC},
consisting of two flat mirrors positioned close together and a curved mirror 
at the opposite end of the cavity, the suppression factor $S_k$ for a TEM mode of 
mode order $k$ can be expressed as \cite{gossler2003}:
\begin{equation}
S_k \approx \sqrt{1+ 4\frac{\mathcal{F}^2}{\pi^2}sin^2 \left[ k \arccos \sqrt{1-\frac{L}{2R}} \right]},
\end{equation}
where $\mathcal{F}$ is the finesse of the ring resonator, $L$ the roundtrip length
and $R$ the radius of curvature of the curved mirror. The GEO\,600 detector 
makes use of two consecutive mode cleaner cavities of 8.0\,m 
and 8.1\,m roundtrip length and finesse values of about 2000, providing an overall 
suppression factors in the range of $3 \times 10^5$ to $2 \times 10^6$ for the 
first four mode orders \cite{gossler2003}.

Similar to mode cleaners at the input of the main interferometers, filter cavities
are used at the output port of a \gw interferometer, so-called \emph{Output Mode Cleaners (OMC)}, 
 to suppress on one hand higher
order modes originating from mirror  distortions of the main optics as well as 
reducing the strength of potential RF modulation sidebands (see the following 
two sections) \cite{Prijatelj10, LIGO-T1000276, VIR-0020A-11}.

\subsection{Length stabilisation of cavities and interferometers} \label{sec:length}

For the mode cleaner cavities to work as required, their length has to be kept
constant to ensure resonance condition for the TEM$_{00}$ mode. Control loops
are used to very accurately position all relevant mirrors at their 
operating points and to suppress any outside disturbances, such as seismic 
motion. The error signals for these control loops are usually generated 
by RF modulation/demodulation techniques derived from the so-called \emph{Pound Drever
Hall (PDH)} technique \cite{PDH}.

\begin{figure}
\begin{center}
\includegraphics[width=0.8\textwidth]{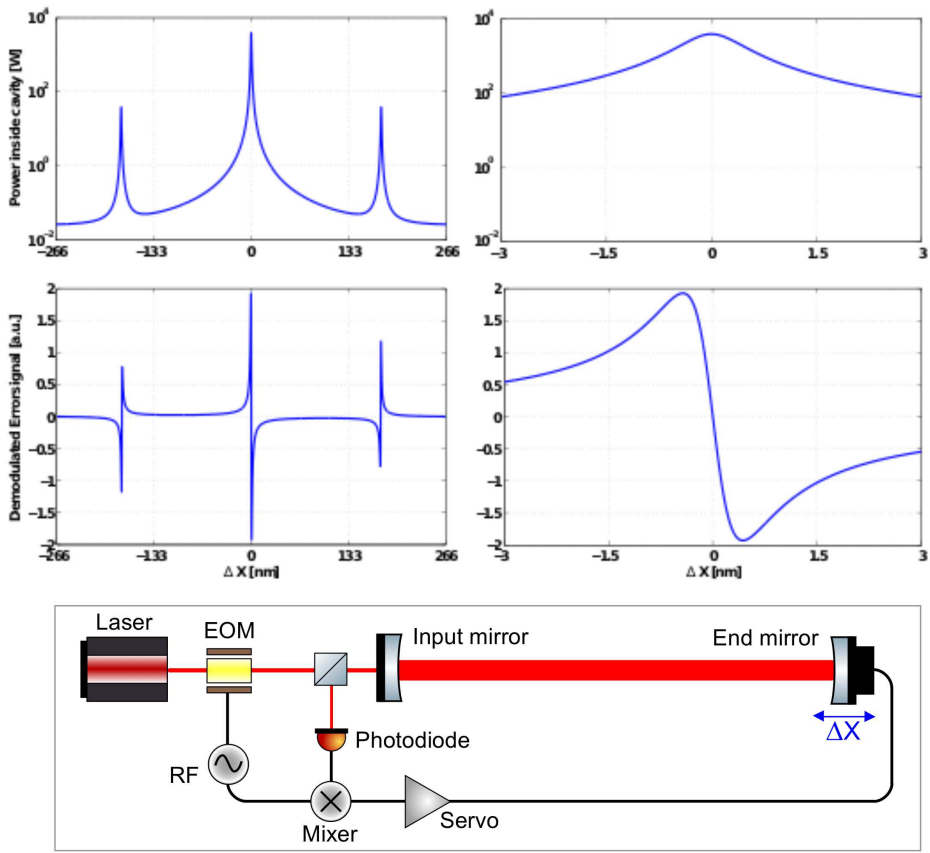}
\caption{Sketch illustrating the principle of Pound Drever Hall technique
used to derive a bipolar error signal for keeping a cavity on resonance for
carrier laser light. Refer to the text for a detailed description.}
\label{fig:PDH}
\end{center}
\end{figure}

Figure \ref{fig:PDH} illustrates the principle of the PDH technique. An 
electro-optic modulator (EOM) is used to imprint RF phase modulation sidebands onto
the laser carrier light. The frequency of these sidebands needs to be chosen
so that the sidebands are not resonant in the cavity and are directly reflected
off the cavity. In contrast, for a cavity length close to resonance, the carrier 
light will at least partly enter the cavity and encounter a phase shift depending
on the microscopic length of the cavity. A photodiode in reflection of the cavity
can then be used to detect the beat of the carrier light leaving the cavity 
and RF sidebands being reflected off the cavity. If the photodiode signal is mixed down 
with the original RF modulation, a bipolar, linear error signal for the cavity length
can be created. The two upper left insets of  Figure~\ref{fig:PDH} show the 
optical power inside the cavity and the demodulated error signal. Three zero-crossings
can be obtained, one for each of the sidebands and the carrier being resonant in 
the cavity. The two upper right insets show  the cavity 
power and the error signal zoomed in around the carrier resonance. The linear range
of the error signal, also referred to as locking range,  in the shown example 
is about a fraction of 
a nanometre.

Modulation/demodulation schemes very similar to the PDH technique can not only
be used to control simple cavities, but also for  all relevant degrees of freedom of \gw
interferometer, such as for instance the differential arm length from which the 
\gw channel is derived. The required length stability of the differential arm length
in advanced \gw detectors is of the order 10$^{-15}$\,m rms and better than 10$^{-20}$\,m$/\sqrt{\rm Hz}$ at audio frequencies. It is worth noting that the PHD technique 
can be extended from longitudinal degrees of freedom to alignment degrees of 
freedom, by using quadrant photodetectors to implement 
so-called \emph{differential wavefront sensing} \cite{ward}.

\subsection{Readout techniques} \label{sec:readout}

\begin{figure}
\begin{center}
\includegraphics[width=1\textwidth]{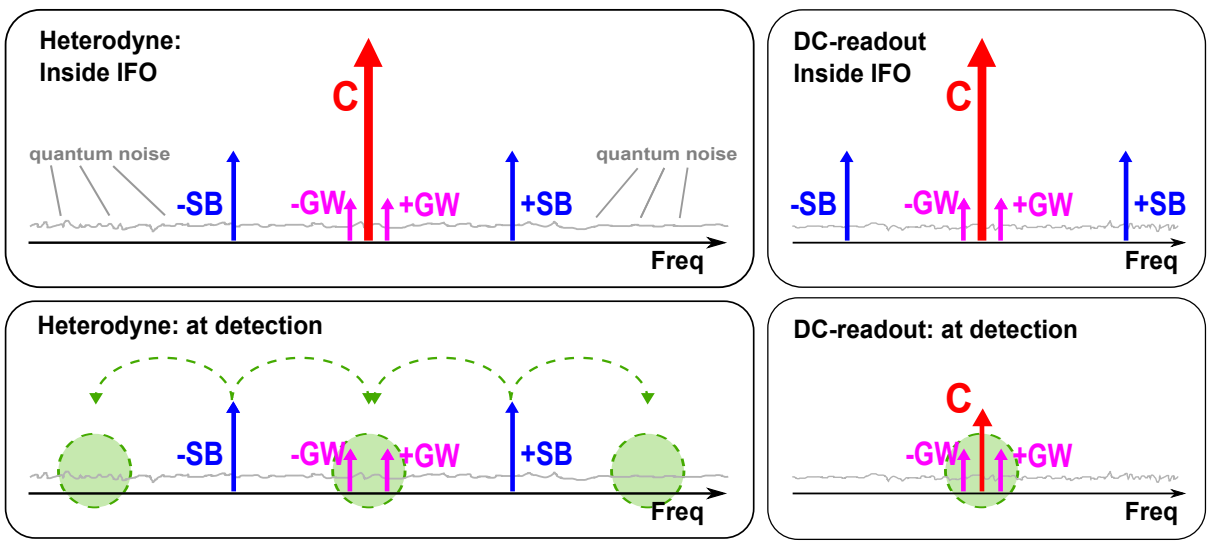}
\caption{Sketch illustrating the principles of different readout schemes. The 
two left hand panels refer to light fields present inside the interferometer (top)
and at the main photodetector (bottom) in the case of heterodyne readout, while 
the two right hand panels illustrate the case of DC-readout. Abbreviations used are 
SB = RF sidebands, GW = GW sidebands (audioband) and C = carrier.}
\label{fig:readout}
\end{center}
\end{figure}

We have already mentioned in the previous section that we can readout 
the differential arm length of a \gw detector by using a RF
modulation/demodulation schemes, often also referred to as heterodyne readout. 
In the following we want to have a closer look at this technique and 
briefly discuss some of the relevant noise terms. The upper 
left box of Figure~\ref{fig:readout} shows the relevant light fields
inside a standard \gw interferometer: the strongest light 
is the carrier (C), followed by the RF phase modulation sidebands
(SB) imprinted in front of the interferometer. If the differential 
armlength is modulated, for instance due to a \gw signal, then phase
modulation signal sidebands (GW) are created in the arms of the interferometer.
The lower
left box of Figure~\ref{fig:readout} shows the relevant light fields
at the output port of the interferometer: at the output port no carrier light is present, because
the interferometer is usually operated at a so-called \emph{dark finge}, i.e.
the differential arm length is chosen to give destructive interference for 
the carrier. However, in contrast to the carrier, the RF sidebands and the 
\gw sidebands interfere constructively at the beam splitter and leave the 
interferometer towards the main photodetector. The photodiode signal is 
then demodulated at the heterodyne frequency to recover an audioband signal 
containing the \gw signal. This demodulation process is indicated by the 
green dashed arrows. The three green circles indicate the frequency 
regions that are beaten down into the \gw channel. Unfortunately only the green circle in the 
center contributes to the signal, while the frequency regions of the outer 
two green circles just contribute additional quantum noise (see Section~\ref{sec:QN}).

Therefore, advanced \gw detectors abandoned heterodyne readout techniques 
and now employ a so-called \emph{DC-readout} \cite{HildDCReadout2009, RW_40m, Fricke}.
The two right hand boxes of Figure~\ref{fig:readout} schematically illustrate the 
DC-readout principle: inside the interferometer exactly the same light 
fields are present as in the heterodyne case. However, at the output port of the 
interferometer there are two major difference in the DC-readout scheme compared to 
the heterodyne case: first of all in DC-readout there is a small amount of 
carrier light present, which is used as local oscillator for the \gw sidebands. 
The presence of the carrier light is achieved by tuning the differential arm length
to be slightly off the dark fringe. The green circle indicates the frequency 
region that is actually sensed. We see that in the case of the DC-readout no
quantum noise from twice the heterodyne frequency is mixed down into the \gw 
channel, which significantly improves the signal-to-quantum-noise ratio
 \cite{Buonanno2003}. So, with DC-readout the heterodyne sidebands do not 
 need to be present at the main photodiode. Therefore, in order to reduce the
 light power on the photodiode and suppress addittional noise couplings, the 
 RF sidebands are filtered out by using an output mode cleaner (see Section~\ref{sec:MC}).
So, in the case of DC-readout the output mode cleaner needs to be designed to
provide both the filtering of higher order TEM modes as well as suppression
of the RF sidebands (which are usually still required for the control of auxiliary
degrees of freedom).
  
\section{Noise Sources in Gravitational Wave Detectors}\label{sec:noise}
There are a number of noise sources which limit the sensitivity of long baseline interferometric detectors in the
frequency range $10$\,Hz-$10$\,kHz. These include fundamental noise sources \cite{fritschel:2007} such as seismic noise,
thermal noise in the mirror test masses, suspensions and coatings, and quantum noise. 

\begin{figure}
\begin{center}
\includegraphics[width=1.0\textwidth]{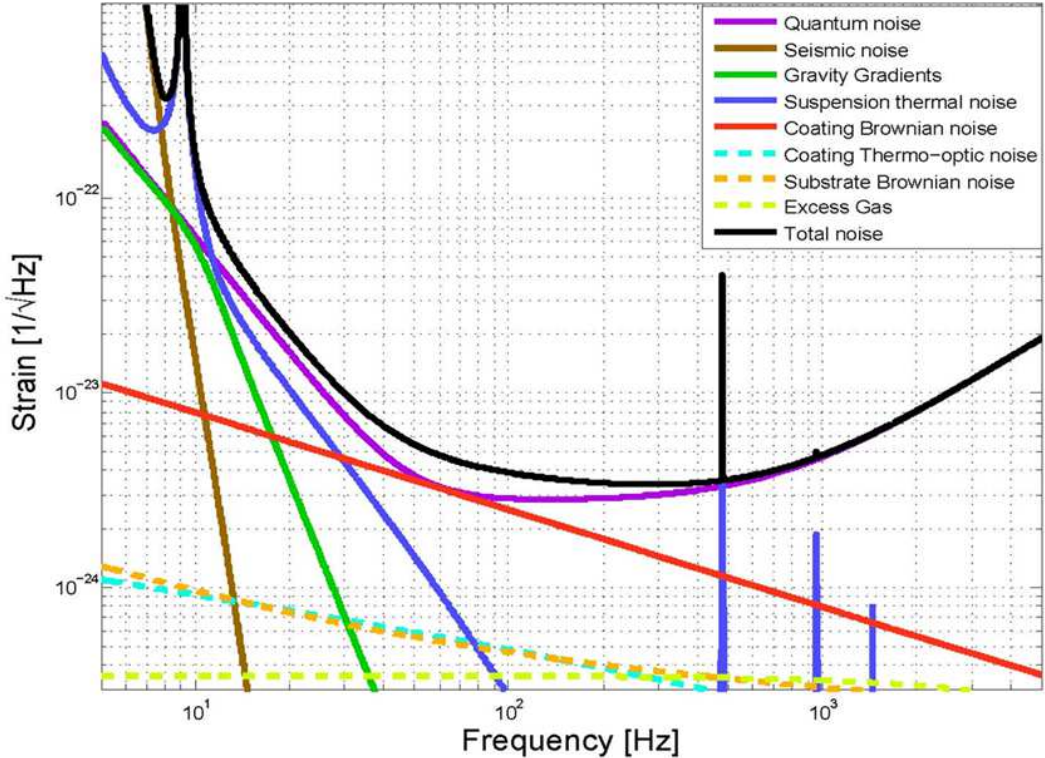}
\caption{The expected limit to performance from the basic noise sources of the aLIGO detector showing the dominant
fundamental noise sources as described in \cite{aLIGO_sens}. This figure was produced using the GWINC software
\cite{GWINC}.}
\label{strain:aLIGO}
\end{center}
\end{figure}
\subsection{Newtonian Noise and Seismic Noise}
In the low frequency regime ($< 10$\,Hz) density perturbations give rise to direct Newtonian couplings to the test
masses of the interferometer. This results in a hard limit to the low frequency performance of detectors as the
gravitational field from these perturbations cannot be shielded. Fluctuations can be either man-made, such as moving
masses, or environmental including density perturbations in the ground due to surface seismic waves. For the advanced
detector network, work is currently underway to assess the feasibility of monitoring the local gravitational field
gradient at the test mass with an array of accelerometers and/or seismometers and performing subtraction from the data
stream either real-time or via post processing \cite{driggers:2012}. For future detectors, either currently under
construction (KAGRA \cite{Somiya:2011}) or proposed (Einstein Telescope \cite{Punturo:2010}), there is the possibility
to locate the detectors a few hundred metres underground where the effect due to surface seismic waves 
is diminished by a factor approximately $10$ \cite{Punturo:2010}.

\begin{figure}
\begin{center}
\includegraphics[width=1.0\textwidth]{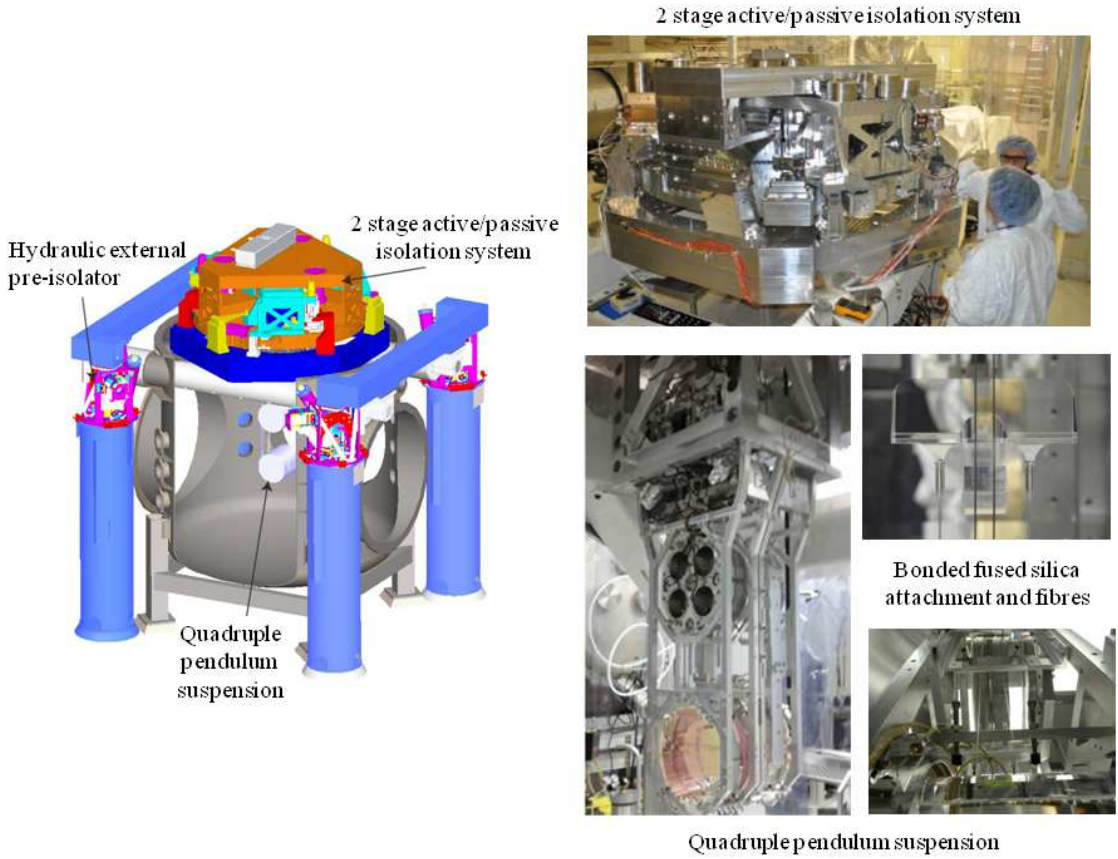}
\caption{Schematic and photographs of the aLIGO Inertial Seismic Isolation system and quadruple pendulum suspension.}
\label{ISI:aLIGO}
\end{center}
\end{figure}

Seismic noise ($< 10$\,Hz) is correlated with periods of stormy weather and human activity. For a quiet site the
acceleration spectral density is approximately $10^{-7}/f^{2}\,\rm{m}\,\rm{s}^{-2}/\sqrt{\textrm{Hz}}$ \cite{Beker:2010,
Peterson:1980}. At frequencies around $0.2$\,Hz the seismic spectrum is dominated by the microseismic peak, which
originates from ocean waves impinging onto the continental plates.  Although this is below the operating bandwidth of
detectors it does contribute to the root mean square of the test mass motion and thus must be reduced in order to bring
the interfermeter mirrors to their operating point. At higher frequencies ($> 10$\,Hz) the seismic noise is mainly
anthropogenic and the seismic spectrum is highly variable between day and night. Seismic noise is minimised by
suspending the test masses of the interferometer from multiple stage seismic isolation systems. These systems are
typically a combination of passive isolation systems and active systems \cite{Acernese:2010, Abbott:2002} to provide
broadband seismic attenuation. It is instructive to consider a mass, $m$, attached to a spring of stiffness, $k$ with
damping constant, $b$. The transfer function between the ground motion, $x_{g}$ and the motion of the mass, $x_{m}$ is
\begin{equation}
\frac{x_{m}}{x_{g}}=\frac{\omega_{0}^{2}}{\sqrt{\left(\omega_{0}^{2}-\omega^{2} \right)^{2} +\omega^{2} \gamma^{2}}},
\end{equation}
where $\omega_{0}=\sqrt{k/m}$ is the resonant angular frequency and $\gamma=\left(b/m\right)$ is the damping constant.
At low frequencies the ground motion and mass motion are identical and the transfer function becomes unity. In effect
the mass and ground move together. At high frequencies, $\omega\gg\omega_{0}$, the transfer function becomes
$\left(\omega_{0}^{2}/\omega^{2} \right)$ (in the limit
of low damping) which provides vibration isolation of the test mass. This is an effective strategy for providing
isolation above a few hertz. In AdV \cite{Acernese:2010} the seismic isolation system comprises multiple
($\simeq7$) stacks of low frequency ($\simeq 0.1\textrm{Hz}$) vertical isolators, and a combination of an inverted
pendulum and pendulum stages for horizontal isolation. For aLIGO \cite{Abbott:2002}, the platform to be isolated
is suspended with stiff springs ($\simeq 5 \textrm{Hz}$) and implemented with seismometers which monitor the platform
motion in all six degrees of freedom. The output of the seismometers is driven to zero via feedback control to
voice-coil actuators, thus providing isolation from the unity gain bandwidth of approximately $0.1$\,Hz up to the point
where passive isolation becomes effective (a few hertz). Figure~\ref{ISI:aLIGO} shows the two stage active/passive
seismic isolation platform utilised in aLIGO which provides a factor of 10 isolation at the microseismic peak
and a factor of 1000 isolation in the $1-10$\,Hz range \cite{Matichard:2010}. The aLIGO detector further
includes an out of vacuum hydraulic actuator which permits large low frequency motion at the level $\pm1$\,mm. This is
an effective technique to take out large low frequency changes in the length of the interferometer arms due for example
to earth tides ($\simeq 600\,\mu{\rm m}$ variation at a frequency of $\simeq 23.1\,\mu{\rm Hz}$).

\begin{figure}
\begin{center}
\includegraphics[width=0.7\textwidth]{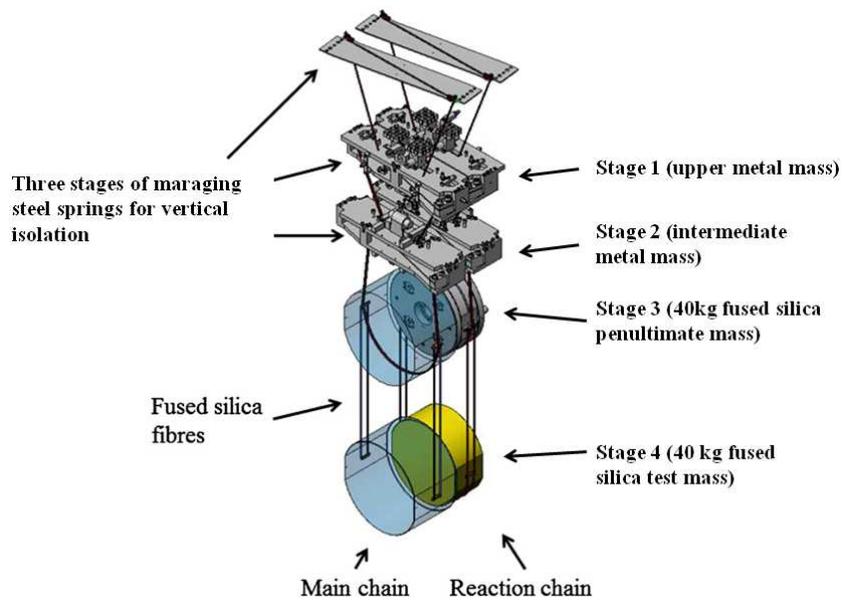}
\caption{Schematic of the aLIGO quadruple pendulum suspension. The $40$\,kg test mass is the lowest
mass in the main chain, while a parallel reaction chain allows for control forces to be applied from
a quiet reference frame.}
\label{quad:aLIGO}
\end{center}
\end{figure}

\subsection{Suspension Thermal Noise} \label {sus_thermal}
The test masses of interferometric gravitational wave detectors are supported by mutiple stage
suspensions that hang below the seismic isolation platforms (see Figures~\ref{ISI:aLIGO} and
~\ref{quad:aLIGO}). The role of the suspension is to further isolate the test mass from seismic
noise in addition to minimising thermal noise. For aLIGO, a quadruple pendulum system
\cite{Aston:2012} is utilised which comprises four pendulum stages for horizontal isolation and
three stages of cantilever springs to provide vertical isolation. The pendulum further allows
the application of control forces for the purpose of alignment and control. These forces are often
termed \textit{local} when they provide damping of a given pendulum suspension or \textit{global}
where the forces are used to control the length and angular degrees of freedom of the interferometer
optical cavities \cite{Aston:2012}. In order not to spoil the seismic attenuation the forces are
applied from a parallel chain called the reaction chain. The control forces are applied in a
hierarchical way \cite{Strain:2012} such that large forces are applied to the upper stages of the
suspension via coil-magnet actuator, and smaller forces (with peak forces in the $\mu$N range) are
applied directly to the test mass via an electrostatic actuator.

To minimize the thermal noise associated with the pendulum modes of the suspension, the final stage
consists of a silica mirror, $40$\,kg in mass (the test mass), suspended from another silica mass (the penultimate mass)
by four silica fibres. These fibres are fabricated using a CO$_{2}$ laser as described in \cite{Heptonstall:2011,
Cumming:2012}. Suspension thermal noise arises due to mechanical dissipation in the materials which make up the
suspension (Brownian noise) or via coupling of statisical temperature fluctuations through the
thermo-mechanical
propoerties of the suspension materials (Thermoelastic noise) and is conveniently calculated via the
Fluctuation-Dissipation theorem \cite{Callen:1951}. Above the pendulum resonance the
displacement thermal noise ($x_{\rm susp}$) due to the suspension is given by
\begin{equation}\label{thermal}
x_{\rm susp}^{2}={\frac{4k_{B}T \omega_{0}^{2} \phi_{\rm total}}{m \omega^{5}}},
\end{equation}
where $T$ is the temperature, $m$ is the pendulum mass, $\phi_{\rm total}$ is the mechanical loss of the pendulum
($\propto
1/Q$ with Quality factor $Q$), $\omega_{0}$ is the resonant angular frequency, $k_{B}$ is Boltzmann's constant and
$\omega$ is the angular frequency. Application of Equation~\ref{thermal} should be applied for all resonant modes of the
suspension, although in principle the pendulum mode ($\simeq 0.6$\,Hz) and the vertical bounce mode ($\simeq 9$\,Hz),
which enters via a 0.1\% cross-coupling, are the dominant terms.

To minimize the displacement thermal noise in room temperature detectors requires the use of ultra-low dissipation
materials with low loss angle such as fused silica. The final stage consists of the fused silica test mass attached to
the fused silica penultimate mass via four fused silica fibres approximately $400\,\mu{\rm m}$ in diameter and $60$\,cm
long \cite{Cumming:2012}. The fibres are welded to silica ears which are attached to the sides of the mass using
hydroxide-catalysis bonding (silicate bonding) \cite{Gwo:2001, Gwo:2003}. The final stage is thus a monolithic fused
silica suspension exhibiting extremely low thermal noise. The dominant contributions to the mechanical loss, and thus
thermal noise performance, will now be briefly discussed. Surface loss \cite{Gretarsson:1999, Penn:2006} originates from
defects on the surface such as dislocations, un-terminated dangling bonds and surface cracks and is an important loss
mechanism for fibres which have a high surface to volume ratio 
\begin{equation}
\phi_{\rm surface}=\frac{4 h \phi_{s}}{r},
\end{equation}
where $h \phi_{s}$ is related to the level of surface damage on the fibre surface and $r$ is the fibre radius.
Thermoelastic loss \cite{Cagnoli:2002} arises from the fact that bending a suspension fibre leads to heating/cooling via
the thermal expansion coefficient. Heat flow across the fibre leads to dissipation. When the fibre is under tension the
variation in Young's modulus with temperature leads to an additional thermoelastic contribution. For fused silica these
two terms have opposite sign and the thermoelastic loss can be cancelled in the bending region by suitable choice of the
fibre geometry \cite{Bell:2013}. Thermoelastic loss is given by
\begin{equation}
\phi_{\rm thermoelastic}=\frac{YT}{\rho C} \left( \alpha-\frac{\sigma \beta}{Y}\right)^{2} \left( \frac{\omega \tau}{1+
\left( \omega \tau \right)^{2}} \right), \label{TEnoise}
\end{equation}
where $Y$ is the Young's modulus of the fibre, $C$ is the specific heat capacity of the material per unit mass, $\rho$
is the density, $\alpha$ is the coefficient of thermal expansion, $\sigma$ is the static stress in the fibre due to the
suspended load, $\beta=\left(1/Y \right) \left(dY/dT \right)$ is the thermal elastic coefficient and $T$ is the
temperature. The characteristic time for the heat flow across the fibre is defined as $\tau$ which for a circular cross
section fibre is \cite{Cagnoli:2002}
\begin{equation}
\tau=\frac{1}{2.16 \pi}\frac{\rho C r}{\kappa},
\end{equation}
with thermal conductivity $\kappa$. For the advanced detectors (aLIGO and AdV) the fibre design is carefully
chosen such that at the bending point of the suspension the tensile stress is chosen to null the thermoelastic noise
contribution for the pendulum mode. The remaining dominant mechanical loss terms are weld loss and bond loss. Weld loss
arises from material which has been heated with a $\textrm{CO}_{2}$ laser \cite{Cumming:2012} to fuse the silica
suspension fibres to the attachment ears on the side of the test mass. This material exhibits a loss which is higher
than the bulk loss \cite{Heptonstall:2010} and likely correlated with the level of thermal stress. Bond loss arises from
the fused silica ears silicate bonded to the side of the test mass. The silicate bonding process produces a strongly
cross linked structure which allows glassy materials to be reproducibly attached in a mechanically and thermally stable
way \cite{Rowan:1998}.

A pendulum mirror suspension stores energy both in the elasticity of the fibre material and the gravitational field. The
latter term is lossless and dominates in heavily loaded suspension fibres. This implies that the pendulum loss is lower
than that of the material used for the suspension fibre. This property is termed dissipation dilution, $D$ and allows
the total mechanical loss to be diluted to $\phi_{\rm total}=\phi/D$. The mechanical loss, $\phi$, is conveniently
calculated via Finite Element methods which evaluate the loss contributions at each point of the suspension (surface
loss, thermoelastic loss, weld loss, bond loss) and scale the loss with the appropriate bending energy stored at that
point in the suspension (e.g. a lossy region with zero stored bending energy will not contribute to the loss)
\cite{Cumming:2009}. The total diluted loss, $\phi_{\rm total}$, is then determined by summing all contributions over
the entire weld/fibre and used to calculate the thermal displacement noise in Equation~\ref{thermal} for all modes of
the suspension stage.

\subsection{Mirror and Coating Thermal Noise}

The mirrors used in the advanced detectors are $40$\,kg to ensure that radiation pressure noise at low frequencies is
sufficiently lowered. The mirrors are cylinders of high quality fused silica with a surfaces roughness of $\simeq
0.16$\,nm rms and a radius of curvature of roughly $2$\,km \cite{Harry:2010}. The aspect ratio of the mirrors is
chosen such that the internal resonant modes are sufficiently high frequency ($\simeq 10$\,kHz and above) to minimise
thermal noise and that flexing of the mirror due to the applied Gaussian laser beam does not introduce significant
coating thermal noise. The resonant modes of the mirror substrate have extremely high quality factors in excess of
$10^{7}$ \cite{Penn:2006}. These high frequency modes store the majority of the $1/2 k_{B}T$ of thermal energy and the
off-resonance thermal noise, in the detector band $10$\,Hz-$10$\,kHz, is sufficiently lowered. For advanced detectors
the mirror substrate thermal noise is significantly lower than the contribution due to the coating thermal noise.

\subsubsection{Mirror Thermal Noise}
Mirror thermal noise originates from mechanical loss in the fused silica substrates and comprises contributions from
surface loss, thermoelastic loss and bulk loss. The origins of surface and thermoelastic loss have been described in the
previous section, while bulk loss originates from the fact that the strained Si-O-Si bonds in fused silica have two
stable minima at different bond angles. Redistribution of the bond angles under thermal fluctuations leads to mechanical
dissipation. The technique proposed by \cite{Levin:1998} is a convenient method to estimate both mirror substrate and
coating thermal noise.  A force, $F_{0}$, is applied onto the front surface of the test mass, in this case a Gaussian
pressure originating from the laser beam, produces a strain energy $\epsilon$ and a dissipated power $W_{\rm diss}=2 \pi
f \int_{V} \epsilon \phi dV$ due to mechanical loss $\phi$. The power spectral density of the thermal displacement noise
is then determined from
\begin{equation}
x_{\rm thermal}^{2}=\frac{2 k_{B} T W_{\rm diss}}{\pi^{2} f^{2} F_{0}}.
\end{equation}
This technique is convenient as it can handle non-homogeneous systems whereby the loss can be varied as a function of
position. This case is particularly useful for modelling both surface loss and bulk loss of the mirror substrate whereby
the appropriate strain energy in different parts of the mirror can be determined. It is however instructive to consider
closed form analytical solutions and for the case of a semi-infinite mirror substrate the thermal noise due to bulk
dissipation (or Brownian dissipation) may be written \cite{Harry:book}
\begin{equation}
x_{\rm mirror}^{2}=\frac{2 k_{B} T}{\sqrt{\pi^{3}} f} \frac{\left(1-\sigma^{2}\right)}{Y w_{m}} \phi_{\rm substrate},
\end{equation}
where $\sigma$ is the Poisson's ratio and $w_{m}$ is the radius of the beam where the electric field has fallen to
$1/e$. Utilising large radius laser beams is clearly an advantage in order to average over a larger surface area. The
mechanical loss of the substrate $\phi_{\rm substrate}$ as determined from a semi-empirical model has been used to
describe the loss contribution in fused silica mirror substrates \cite{Penn:2006}
\begin{equation}
\phi_{\rm substrate}=C_{1} \left(\frac{S}{V} \right)+ C_{2} \left(\frac{f}{1\,{\rm Hz}} \right)^{C_{3}},
\end{equation}
where $V$ is the sample volume, $S$ is the surface area and $f$ is the frequency. The coefficients $C_{1}$ and $C_{2}$
and exponent $C_{3}$ relate to the specific type of fused silica and typical values are $C_{1}= (6.5 \pm 0.2) \times
10^{-9}$,
$C_{2}= (6.3 \pm 0.2) \times 10^{-12}$, $C_{3}= 0.77 \pm 0.02$ \cite{Penn:2006}. An analytical form for the
thermoelastic noise, which couples via the thermal expansion coefficient $\alpha$, may also be determined
\begin{equation}
x_{\rm thermoelastic}^{2}=\frac{4 \left(1+\sigma^{2} \right) \kappa \alpha^{2} k_{B} T}{\pi^{2.5} C^{2} \rho^{2} w_{m}^{3} f^{2}}.
\end{equation}
Corrections which take into account finite sized mirrors \cite{Bondu:1998} are also readily calculated although for the
advanced detector mirror substrates corrections, $C_{\rm fsm}$, are $\simeq 0.98$. Again utilising a large radius laser
beams is clearly advantageous to average over a larger surface area of the mirror. For the advanced detectors the
contribution due to the mirror thermoeasltic noise is negligible and thus not shown in Figure~\ref{strain:aLIGO}.

\subsubsection{Coating Thermal Noise}
Coating thermal noise originates from the multilayer dielectric coating stacks which make up the high reflectivity
coatings on the test masses. The noise contribution can be reduced by utilising low mechanical loss coatings and by
averaging over a larger surface of the mirror by utilising larger laser beam diameters. For advanced detectors a beam
radius (1/e) of the order $\simeq 5-6$\,cm is used on the mirrors as a compromise between lowering coating thermal noise
and reducing clipping losses to below 1\,ppm and cavity instabilities \cite{Harry:2006}. The
coating layers
are made up of alternating stacks of silicon dioxide (SiO$_{2}$), or silica, as the low refractive index layer and
tantalum pentoxide (Ta$_{2}$O$_{5}$), or tantala, as the high refractive index layer. The tantala/silica coating
material exhibits mechanical losses $3\times 10^{-4}$/$5\times 10^{-5}$ respectively \cite{Harry:book, Harry:2006a}
which is somewhat higher than the fused silica used in the mirror substrates and suspensions. This
results in coating thermal noise being an important contribution in the frequency around $75-100$\,Hz. Significant
progress has been made in lowering the thermal noise in coating layers via the additional of dopants including titanium
dioxide (TiO$_{2}$), or titania, in the tanatala coatings \cite{Harry:2006b}. These dopants at the level of $\simeq
25\%$ have been shown to lower the mechanical loss by approximately 40\%. There has also been significant progress in
understanding the link between mechanical loss and coating microstructure \cite{Bassiri:2011} and this is an evolving
area in which the fields of gravitational wave detection and solid state physics are efficiently collaborating to
provide low loss high reflectivity mirror coatings. This work also has significant applications in the field of
precision metrology where low thermal noise mirror coatings are necessary. Expressions for the Brownian and
thermoelastic coating thermal noise can be derived for multilayer dielectric coatings to yield for semi-infinite
substrates \cite{Harry:book}
\begin{equation}
x_{\rm coating}^{2}=\frac{2 k_{B} T \phi_{\rm eff} \left(1-\sigma \right)}{\pi^{3/2}f w_{m} Y}, \label{CTN}
\end{equation}
with
\begin{equation}\label{coating_loss_angle}
\phi_{\rm eff}=\phi+\frac{d}{w_{m} \sqrt{\pi}}\left(\frac{Y}{Y_{\perp}}\phi_{\perp}+\frac{Y_{\parallel}}{Y
\phi_{\parallel}}
\right),
\end{equation}
where $\phi$ and $Y$ denote the mechanical loss and Young's Modulus of the substrate, and the subscripts $\perp$ and
$\parallel$ denote the perpendicular and parallel components of the coating parameters respectively. 

Thermo-optic noise in the coating is treated as contribution from thermoelastic noise due to deformation of the coating
surface and thermorefractive noise arising from variations in the refractive index of the coating with temperature
fluctuations \cite{Evans:2008}. This effect is analogous to the suspension thermoleastic noise which can be nulled by
suitable choice of the fibre dimension as described in section \ref{sus_thermal}. For advanced detector coatings the
thermorefractive noise can be written
\begin{equation}
x_{\rm thermoelastic}^{2}=\frac{2\sqrt{2} k_{B} T^{2} }{\pi^{3/2} w_{m}^{2} \sqrt{\kappa C f}} \left(C_{\rm fsm} \alpha d
-\beta \lambda \right),
\end{equation}
where symbol definitions are identical to section \ref{sus_thermal}, $C_{\rm fsm}$ is the finite mirror correction
factor ($\simeq 0.98$), $d$ is the coating thickness, $\beta=dn/dT$ is the variation in refractive index with
temperature and $\lambda$ is the wavelength. 

The advanced detectors will utilise optimised coating thicknesses that are slightly different to the standard
$\lambda/4$ dielectric stacks \cite{Villar:2010}. This results from the fact that it is the combination of mechanical
loss multiplied by coating thickness that determines the level of thermal noise contribution
(Equation~\ref{coating_loss_angle}).  As the losses of the tantala and silica coatings are different, an optimum can be
achieved that is slightly different to a standard quarter wave layer thickness. The reader is referred to
Figure~\ref{strain:aLIGO} where the dominant noise terms are plotted as a function of frequency.

\subsection{Quantum Noise}
\label{sec:QN}

Quantum noise plays a very special role in gravitational wave interferometers. First of all, as can be seen in
Figure~\ref{strain:aLIGO}, quantum noise is the limiting fundamental noise source over most frequencies in the aLIGO
detection band, and therefore it is of ultimate importance to further improve it. Secondly, as indicated by its name
quantum noise directly arises from the quantum nature of photons. Two different noise mechanisms contribute to quantum
noise: 
\begin{itemize}
 \item \textbf{Photon shot noise:} Due to the fact that the photons in a laser beam are not equally distributed in time, 
 but follow a Poisson distribution, any laser beam detected by a photo diode will cause the output of the photo diode
 to carry noise. This shot noise, which scales proportional to the square root of the detected optical power, can be 
 considered as the readout or sensing noise of the interferometer.  
 \item \textbf{Quantum radiation pressure noise:} When the laser beam is reflected by a mirror, the photons transfer
 momentum onto the mirror, or in other words there is radiation pressure force acting on the mirror. Since, as stated
 above, the  photons are not equally distributed in time, the radiation pressure force on the mirror fluctuates.
 However, due to the fact that the mirrors in gravitational wave detectors are suspended from a chain of pendulums, the
 mirrors are susceptible to the radiation pressure force fluctuations and as a consequence the mirror position
 fluctuates. This  quantum radiation pressure noise can be considered as a form of back-action noise from the
 measurement process itself. 
\end{itemize}
   
It is interesting to compare how these two quantum noise components scale for different optical power. We have 
mentioned that the shot noise detected by a photo diode increases with the square root of the optical power. However,
because the actual gravitational wave signal scales linearly with the optical power, overall we can win in the \gw
signal to shot noise ratio by increasing the power, $P$. The amplitude spectral density of shot noise equivalent strain
follows:
\begin{equation}
h_{\rm{sn}} \propto \frac{1}{\sqrt{P}}\;.
\end{equation}
In contrast, the contribution of quantum radiation pressure noise increases linearly with the optical power. The
amplitude spectral density of radiation pressure noise equivalent strain is given by: 
\begin{equation}
h_{\rm{sn}} \propto \frac{P}{m f^2}\;,
\label{eqn:RPN}
\end{equation}
where $m$ is the mass of the mirror and $f$ is the frequency. 

The opposite scaling of the two quantum noise components with respect to the optical power circulating in the optical 
system, gives rise to the so-called standard quantum limit (SQL) \cite{BraginskyRMP1996}, which can be interpreted to 
be the equivalent of the Heisenberg Uncertainty Principle applied to interferometry of uncoupled test masses.  The
higher the optical  power, the lower the shot noise contribution and therefore the better our readout accuracy, while 
at the same time we increase the back action noise and therefore disturb the mirror positions. The SQL poses a
fundamental sensitivity limit for any classical interferometer. However, as we will discuss below, there are various 
non-classical techniques available which potentially allow one to surpass the standard quantum limit.

\begin{figure}
\begin{center}
\includegraphics[width=0.7\textwidth]{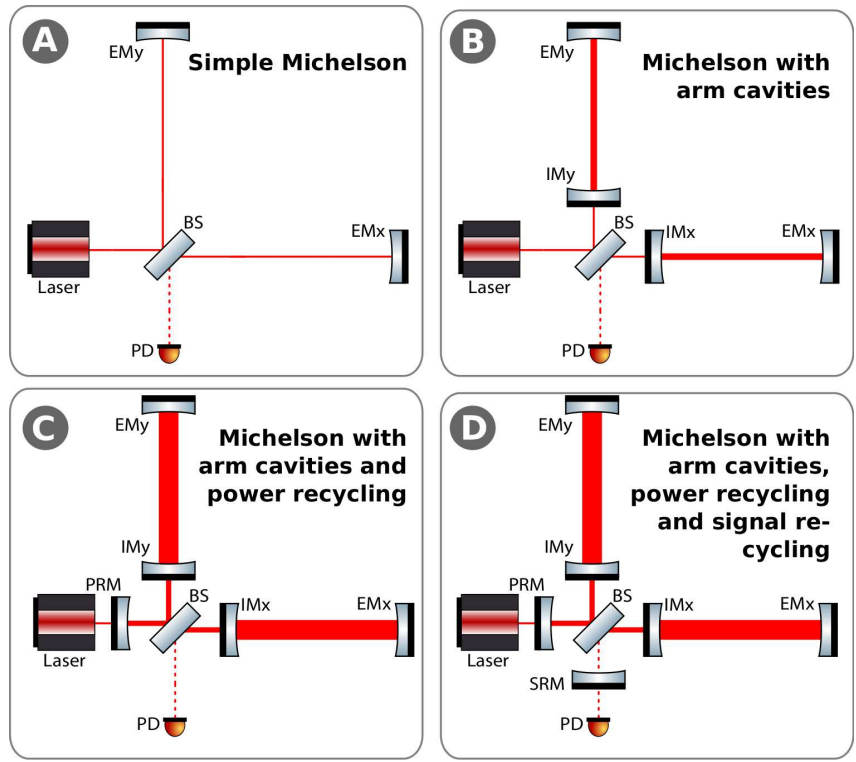}
\caption{Evolution of interferometer layouts from a simple Michelson (A) to advanced gravitational wave detectors as
advanced LIGO (D). The following abbreviations have been used: BS = beam splitter, IM = input mirror, EM = end mirror,
PD = photo detector, PRM = power recycling mirror and SRM = signal recycling mirror.}
\label{fig:ifo_configs}
\end{center}
\end{figure}

For the moment let us go back to aLIGO and discuss its interferometric configuration, as shown in the lower right panel
of Figure~\ref{fig:ifo_configs}. Clearly the aLIGO optical layout is more complicated than the simple Michelson
interferometer shown Figure~\ref{fig:ifo_GW}, and this is what we should understand.

\begin{figure}
\begin{center}
\includegraphics[width=0.8\textwidth]{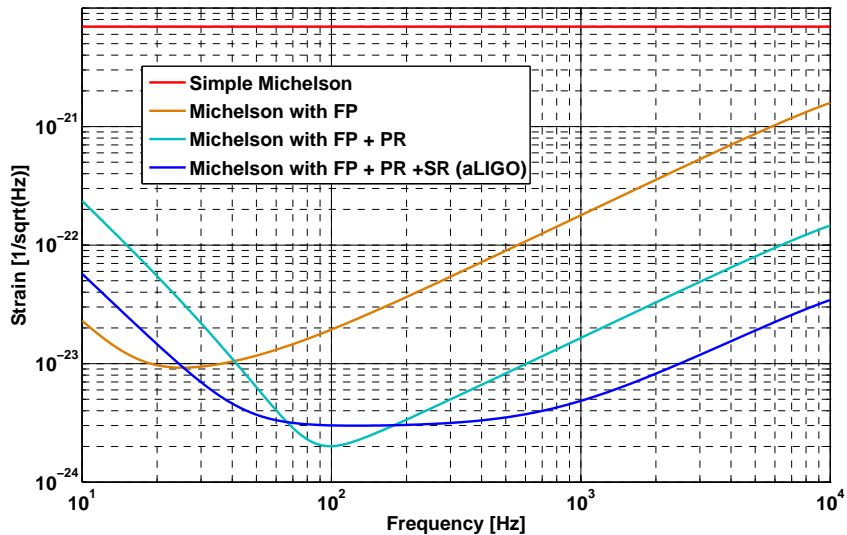}
\caption{Quantum noise limited strain sensitivities for the four interferometer configurations displayed in Figure
\ref{fig:ifo_configs}}
\label{fig:ifo_configs_strain}
\end{center}
\end{figure}

Let us start from a simple Michelson interferometer with 4\,km long arms (see panel A of Figure~\ref{fig:ifo_configs})
and a laser that the delivers 125\,W of input power \cite{Kwee:12}. It is most practical to stabilise (so-called
\emph{locking}) the differential arm length of the interferometer so that  the beams returning from the two arms
interfere completely  destructive at the beam splitter. Therefore, in normal operation and in absence of any strong
gravitational waves there will be no light at the output port and on the photo diode. Only  when there is a change in
the differential arm length of the Michelson (maybe due to the presence of a sufficiently strong gravitational wave, or
due to an external disturbance acting on the mirrors) the destructive interference will be partly destroyed  and light
emerges at the output of the interferometer. The corresponding quantum noise limited strain sensitivity of such a simple
Michelson interferometer  is shown as the red trace in Figure~\ref{fig:ifo_configs_strain}; such an instrument would be
limited entirely by shot noise and would provide a flat spectrum at a level of $6 \times 10^{-21}/\sqrt{\rm Hz}$.
 
In order to increase the interaction time of the laser light with a potential gravitational wave most advanced
gravitational wave detectors employ Fabry-Perot cavities in the arms to resonantly enhance the light power in the arms
(see panel B of Figure~\ref{fig:ifo_configs}). In the case of aLIGO these arm cavities feature a finesse of about 400.
The resulting sensitivity improvement can be see in the orange trace in Figure~\ref{fig:ifo_configs_strain}. At the
sweet spot of 20\,Hz the quantum noise limited sensitivity improves by about a factor of 600 compared to a simple
Michelson interferometer. It is also obvious that the sensitivity obtained with arm cavities is not flat any more; above
30\,Hz the sensitivity rolls off due to the presence of the arm cavity pole. In addition, using the arm cavities we have
increased the laser power so strongly that below 25\,Hz quantum radiation pressure noise starts to dominate quantum
noise.   

In a next step to further enhance the light power inside the interferometer, we apply a technique called power recycling
\cite{PowerRecyclingSchnier1997} (see panel C of Figure~\ref{fig:ifo_configs}). As we have discussed above, when set to
destructive interference no light is leaving the interferometer towards the output port. For energy conservation reasons
this means that all the light returning from the interferometer arms interferes constructively into the input port,
i.e.\ the light goes back towards the laser. This light which would be wasted otherwise can be recycled by placing a
semi-transparent mirror, the power recycling mirror (PRM), between the laser and the beam splitter. The effect of power
recycling on the quantum noise limited strain sensitivity is displayed in the cyan curve of
Figure~\ref{fig:ifo_configs_strain}. Due to increasing the circulating light power in the interferometer arms to 0.8\,MW
we can significantly improve the sensitivity above 100\,Hz. However, since we have increased the contribution of
quantum radiation pressure noise at low frequencies by the same amount we have reduced shot noise at high frequency, we
overall lose sensitivity below 40\,Hz. 

Finally, all advanced \gw detectors will also make use of a technology called signal recycling \cite{Meers1988}. 
Similarly, to how power recycling recycles the main carrier light leaving towards the input port, by inserting a signal
recycling mirror between the beam splitter and the output port (see panel D of Figure~\ref{fig:ifo_configs}), we can
recycle  the signal sidebands leaving the interferometer towards the output port. aLIGO employs a specific flavour  of
signal recycling called resonant sideband extraction. The aim here is to widen the arm cavity bandwidth for the signal
sidebands and therefore to more easily extract the side bands by forming a coupled three mirror cavity of SRM, IM and
EM, to provide a reduced  effective cavity finesse for the signal sidebands.  The effect of this
technique applied to the
aLIGO configuration can be seen in  the blue trace of Figure~\ref{fig:ifo_configs_strain}. Using signal recycling
allows us to significantly widen the instrument response by improving the sensitivity on the low and at the same time
high frequency end of the detection band. The application of signal recycling is planned for all future gravitational
wave detectors currently in construction or under consideration.

In order to further improve the quantum noise limited strain sensitivity of future gravitational wave detectors, and
eventually surpass the SQL, three different concepts have been suggested:
\begin{itemize}
\item The injection of squeezed vacuum states
\item Techniques exploiting opto-mechanical rigidity
\item Speedmeter configurations
\end{itemize} 

\subsubsection{Squeezed Vacuum}

The presence of quantum noise in a laser-interferometric \gw detector can also be understood by vacuum fluctuations of
the electrical field entering the interferometer through any open port. For instance the output port of the
interferometer can be considered as an open port which allows the vacuum fluctuations to enter the \gw detector, interact
with its mirrors (back-action noise) and finally be detected on the main photo detector (sensing noise).  

\begin{figure}
\begin{center}
\includegraphics[width=0.6\textwidth]{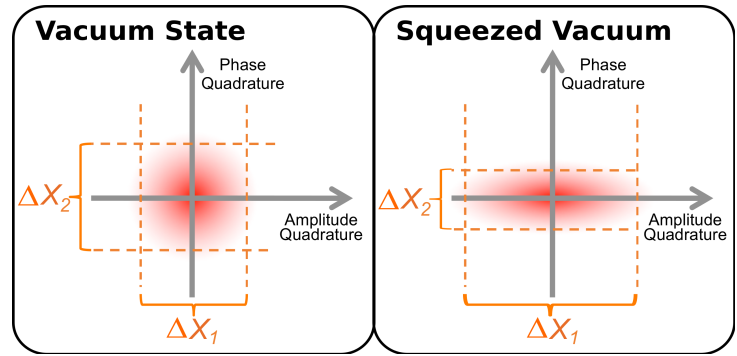}
\caption{Quantum noise can also be described as vacuum fluctuations. The left
panel shows a classical vacuum state with a circular probability distribution.
A squeezed vacuum state (right panel) can be used to reduce the vacuum 
noise one readout quadrature, while increasing the noise present in the other
quadrature.  }
\label{fig:squeez_vac}
\end{center}
\end{figure}

As displayed in the left hand panel of Figure~\ref{fig:squeez_vac} the vacuum fluctuations feature a circular shape in
the quadrature picture. It has to  be noted that for a pure vacuum state the noise in the amplitude and phase
quadratures is uncorrelated. The radiation pressure noise measured in the \gw detector scales with the noise in the
amplitude quadrature, while the shot noise component of the quantum noise scales with the noise in phase quadrature.
Therefore it is obvious that we can redistribute or shape the spectrum of the quantum noise limited sensitivity by
changing the shape of the vacuum fluctuations. 

While the Heisenberg Uncertainty Principle forbids us to reduce the overall area of this circle, the application of
squeezing \cite{Caves1981} allows us to  manipulate the shape of the vacuum noise and to transform the circle into a 
squeezed ellipse (see right panel of Figure~\ref{fig:squeez_vac}). 
Experimentally, such states can be realised using non-linear optical technologies.
In a first step some 1064\,nm light is taken off the main laser and is converted
into frequency doubled light with a wavelength of 532\,nm using a second harmonic 
generation (SHG). This light is then fed into an optical parametric oscillator (OPO),
where parametric down conversion is used to created a correlated pair of 1064\,nm
photons from a single 532\,nm photon. If the OPO is operated below threshold, squeezed 
vacuum states are created \cite{McClell}.

The application of squeezed light to enhance the sensitivity of \gw observatories has been matured over the past decade.
The GEO\,600 interferometer makes regular use of the injection of squeezed light states since the year 2010
\cite{Geo_squeez} and was used to demonstrate long-term application of squeezing as well as for long duration noise
studies \cite{Grote2013}. Recently also in one of the LIGO interferometers squeezing tests have been performed to
demonstrate a sensitivity improvement at the mid to low-frequency end of the detection band \cite{Aasi2013c}. 

\begin{figure}
\begin{center}
\includegraphics[width=0.9\textwidth]{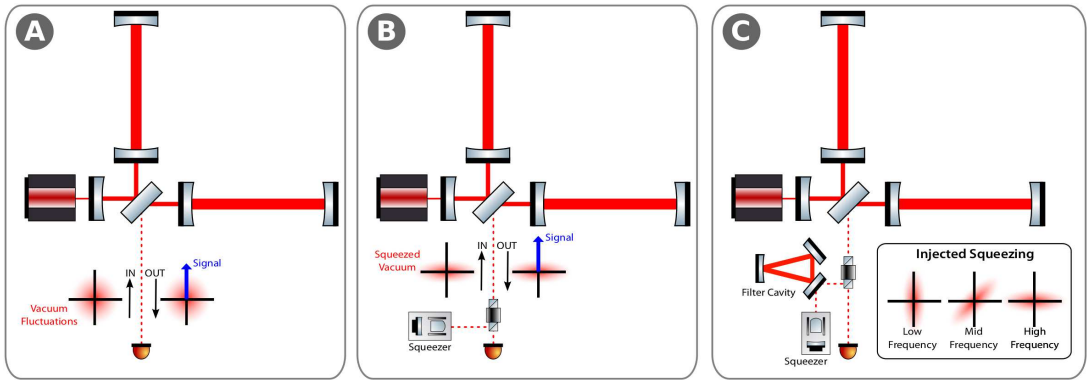}
\caption{Illustrations of a \gw detector without any squeezing (A), the injection of phase squeezed
vacuum states (B) and the injection of squeezed vacuum states 
with a frequency dependent squeezing angle (C), which can be obtained by using
the dispersion of a filter cavity from which the squeezed light is reflected. }
\label{fig:Squeez_injec}
\end{center}
\end{figure}

Figure \ref{fig:Squeez_injec} schematically illustrates the injection
of squeezed states into a \gw detector. Phase squeezed vacuum is created by
the squeezing source and injected via a Faraday isolator at the output port
of the interferometer. The squeezed vacuum state is reflected from the interferometer
and, after passing again through the Faraday isolator, is detected together 
with any potential \gw signal on the main photodetector. A comparison of the 
output fields in the left and center panels of Figure~\ref{fig:Squeez_injec} 
show an increased signal (blue arrow) to noise (red ball/ellipse) ratio for the 
injection of squeezed light.

\subsubsection{Opto-mechanical Rigidity}

The high light powers employed by gravitational wave detectors create significant radiation pressure effects. Forces
acting on the test masses originating from light pressure can be of the same order or even larger than mechanical
restoring forces acting on the test mass. Therefore such interferometers are described as \emph{opto-mechanical} systems
rather than just mechanical or optical systems. For instance in a cavity slightly off resonance the circulating light
power strongly depends on the relative position of the cavity mirrors. For small changes  the cavity power and therefore
also the force acting on the mirrors is linearly proportional to the relative position of the mirrors. The resulting
opto-mechanical behavior resembles spring characteristics similar to Hooke's law (but with negative damping).

Such optical springs \cite{Corbitt2006, Corbitt2007} can cause various types of challenges and instability problems 
for lock-acquisition and controlling an advanced gravitational wave detector (for instance Sidle-Sigg instability
\cite{sidles-sigg}). However, on the other hand optical springs can also be utilised to further improve the quantum 
noise limited sensitivity of interferometers and even to surpass the SQL. 

\begin{figure}
\begin{center}
\includegraphics[width=0.8\textwidth]{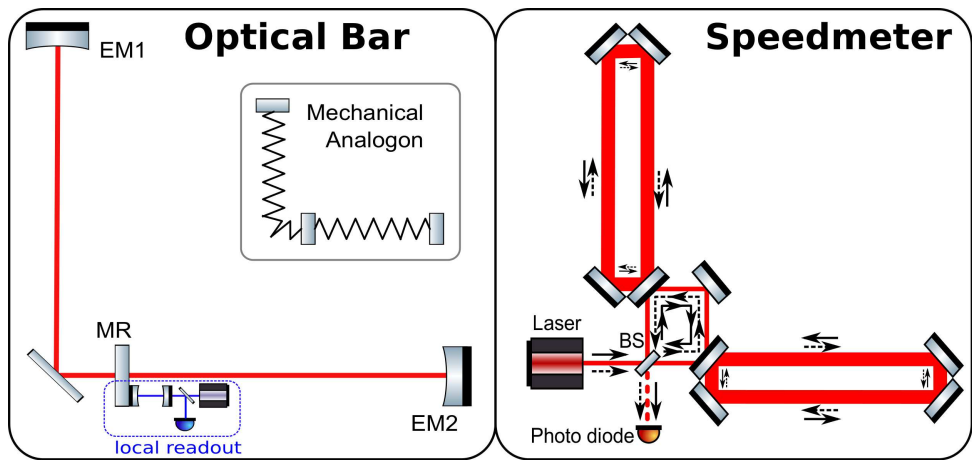}
\caption{Schematic layouts of an Optical Bar configurations (left) and a Sagnac interferometer based speedmeter
configuration (right).}
\label{fig:bar_speed}
\end{center}
\end{figure}

The left panel of Figure~\ref{fig:bar_speed} shows the simplest example of an optical spring enhanced gravitational wave
detector, a so-called \emph{optical bar} \cite{Braginsky1996}. The central mirror MR is connected to the two end mirrors
(EM1 and EM2) via optical springs. In case of a gravitational wave of plus polarisation incident perpendicular
to plane of the detector, the distance between MR and one of the end mirrors will be stretched so that the spring in
this arm will pull MR, while the distance between MR and the other end mirror will be shortened so that the spring in
this arm will push MR. Hence the optical bar acts as an transducer, which 'converts' the gravitational wave into a
actual movement of MR in respect to its local frame. The gravitational wave signal can then 
be read out by an independent local readout. The optical 
bar scheme actually provides two significant benefits. First of all the test 
masses are converted from free-falling objects into oscillators, providing 
increased response for signals at the optical spring resonance. Secondly, separating
the \gw transducer from the readout provides us with the possibility to
 individually optimise the light powers in each of the systems and therefore, if well designed,
 providing sensitivities below the SQL.

More complex configurations employing optical springs include optical lever \cite{Khalili2002, danilishin2006} and
multiple-optical spring configurations (see for instance \cite{Rehbein2007, Rehbein2008, PhysRevD.84.062002}).     

\subsubsection{Speedmeter Topologies}

The SQL arises during the measurement of our test masses because the test mass positions at different times ($x(t_1)$
and $x(t_2)$) do not commute, i.e.\,
\begin{equation}
[x(t_1), x(t_2)] \neq 0
\end{equation}

This limitation is imposed by quantum mechanics as soon as we measure the absolute position of mirrors (at DC
frequency). However, for \gw detection on the Earth we are anyway only interested at frequencies above about 1\,Hz, so
that in principle we do not necessarily need to measure absolute positions of mirrors to very high precision. 

For example if we were to measure the speed (or better the momentum, $p$) of an ensemble of test masses and made sure to
not read out any absolute position information, we could in principle beat the SQL over a very wide frequency range. In
simple cases momentum can be considered as a conserved observable and hence we can measure the momentum of an ensemble
of test masses  continuously with much higher precision than the test mass positions:
\begin{equation}
[p(t_1), p(t_2)] = 0
\end{equation} 
Knowing the speed of the test masses still allows to reconstruct the gravitational wave signal, while such a measurement
would in first order not be limited by the SQL. This concept, coined \emph{speedmeter}, had first been proposed in the
context of readout of  traditional mechanical bar detectors \cite{Braginsky1990251}, before it was realised that a
Michelson interferometer can be transformed to a speedmeter by adding a so-called \emph{sloshing cavity} at its output
\cite{Purdue2002a}. In 2003 a theoretical analysis by Chen \cite{Chen2003} identified the intrinsic speedmeter
properties of a Sagnac interferometer.
 
The right panel of Figure~\ref{fig:bar_speed} shows the optical layout of a Sagnac speedmeter suitable for gravitational
wave detection. The key difference between a Michelson interferometer and Sagnac interferometer is that, while in the
Michelson all photons only enter one of the two interferometer arms, in the Sagnac speedmeter the photons travel 
through both arms. Therefore each mirror is sensed at two different times in quick succession, once by the beam
transmitted through the beam splitter (BS) and once by the beam reflected from BS. This is the reason why only the
velocity or momentum information of the test masses is present on the light reaching the main photo-detector. 

While the speedmeter concept looks very promising from a theoretical point of view, an experimental demonstration is
missing so far. There are currently efforts to set up a speedmeter proof of concept demonstration experiment \cite{SSE}.

\section{Future sensitivity improvements}
\label{sec:upgrades}
There is significant effort underway in order to install and commission the hardware that is necessary to realise the
sensitivity improvement of the advanced detector network. It is planned that these detectors will come online in 2015
and reach full design sensitivity around 2019 for aLIGO and 2021 for AdV \cite{Aasi:2013}. While this work is
progressing it is essential to consider upgrade scenarios for the advanced detector network which could provide a
further improvement in a factor of 3--5 in the strain sensitivity. As hardware upgrades typically take 10--15 years to
develop from the lab and install into the kilometre scale detectors, there is a strong targeted programme of future R\&D
which is directed at upgrades and future detectors. In this paper we will only discuss future upgrades to the advanced
detector network. However, the reader is referred to the design study of the Einstein Telescope which focusses on a
future 3rd generation detector in Europe. This instrument features underground operation, arm lengths of 10\,km and the
use of dual temperature detector operation to allow reduction of fundamental noise sources \cite{Punturo:2010, ET-D}.

\subsection{Injection of squeezed light with a frequency dependent squeezing angle}

For improving the aLIGO sensitivity  initially in a certain frequency band and later over the full frequency range, the 
application of squeezing has been suggested \cite{Strawman, Red:2012} since of all concepts for improving quantum noise 
beyond the SQL it has achieved the highest maturity.

While during the previous decade the focus of squeezed light research within the gravitational wave community was set on
prototyping suitable squeezed light sources providing sufficient squeezing level (more than 12\,dB) \cite{Mehmet:11} as
well as extending the range of squeezing to frequencies as low as 1\,Hz \cite{Vahlbruch2007}, current research efforts
concentrate the low-noise implementation of squeezed light in a full-scale advanced gravitational wave detector. 
 
In order for squeezing to providing improved quantum noise at all frequencies, i.e. in the frequency range limited by
shot noise as well as in the range limited by back-action noise, squeezed light states need to be injected, where the
squeezing ellipse features a frequency dependent orientation \cite{PhysRevA.71.013806}. Such a frequency dependent
squeezing angle can be realised by using the dispersion in reflection of a detuned cavity, often referred to as filter
cavity (see Figure~\ref{fig:Squeez_injec}).
 
The key difficulty in the realisation of such a filter cavity is to provide a very low bandwidth (of the order of the
frequency cross over between shot and back-action noise) combined with extremely low optical losses in order to not
destroy the squeezing \cite{PhysRevD.88.022002}. 
 
The quantum noise curve for the aLIGO upgrade shown in Figure~\ref{red_warm} assumes a 300\,m long filter cavity with a 
roundtrip loss of 30\,ppm, an input mirror transmission 425\,ppm and a detuning frequency of -16.8\,Hz \cite{Red:2012}.
In addition to further improve the sensitivity at the low frequency end, the weight of the test masses was assumed to be
increased from 40\,kg to 160\,kg.  

\subsection{Warm upgrades}

\subsubsection{Suspension thermal noise }
As noted previously, lowering the suspension thermal noise requires the use of ultra low loss materials. Warm upgrades
(e.g.\ room temperature operation) will likely require the use of heavier test masses ($\simeq 160$\,kg) to lower the
effect of radiation pressure noise (see Equation~\ref{eqn:RPN}). Fortunately, fused silica is available up to diameters
of approximately 55\,cm which allows sufficiently heavy test masses to be fabricated. Suspending heavier test masses
will require the fused silica fibres to be scaled in dimension such that the region of thermoelastic cancellation is
equal to 1.3\,mm (see section \ref{sus_thermal} and Equation~\ref{TEnoise}). With current CO$_{2}$ laser pulling
technology this does not seem to be a challenging requirement. It is also advisable to use longer suspensions as this
increases the ratio of the energy stored in gravity to that stored in elasticity. This has the effect of improving the
dissipation dilution ($D$) and thus lowering the effective mechanical loss angle of the suspension. In aLIGO suspension
lengths of up to 1.5\,m can be accommodated with some re-working of the support structure necessary to hold the seismic
isolation system. It should be noted that it is desirable to maintain the violin mode frequencies of the tensioned fused
silica fibres in the region of 500\,Hz and thus any length increase of the suspension will tend to reduce these
frequencies. One possibility to further increase the frequency is to utilise thinner fibres which operate at a higher
stress. In the advanced detectors the fibres operate at about 16\% of their ultimate tensile strength. Working at 30\%
would ensure that the violin modes remain at the same frequency \cite{Heptonstall:2014}. An added benefit of both a
longer final stage and thinner fibres is that the vertical mode of the suspension lowers from
9\,Hz to 6.2\,Hz, thus
providing additional vertical isolation at 10\,Hz (Equation~\ref{thermal}). Optimising the geometry of the neck of the
fused silica fibres is also a technique to further improve the suspension thermal noise. By pulling from thicker (5\,mm
diameter) fused silica stock more of the energy can be stored in the thin section of the fibre which ultimately
maintains a high dissipation dilution. 

With such improvements to the fused silica final suspension stage, care must be taken to ensure that the performance is
not limited by other components of the quadruple pendulum. Modelling has shown that the final stage of vertical
cantilever springs, which are currently fabricated from maraging steel, will begin to limit the performance of the final
pendulum stage if improvements at the level of 2-3 can be made in the pendulum thermal noise. One
possible solution is
to utilise fused silica springs in this final stage which will have a loss angle roughly $10^{-3}$ lower than the
maraging steel \cite{Hammond:2012}. Challenges which need to be overcome include (i) making springs sufficiently strong
with tensile strength of up to 800\,MPa, and (ii) providing protective overcoats such that the silica springs can be
handled and clamped into the suspension. There is significant interest in the use of coatings such as Diamond Like
Carbon (DLC) to provide robust protective layers.

\subsubsection{Coating thermal noise }
For coating thermal noise, the heavier test mass (with diameter $\simeq 55$\,cm) will allow a reduction of 1.6 in the
coating thermal noise via larger beam diameter. Further improvement will be achieved through the combination of improved
coatings and/or new materials. As mentioned previously, there is strong evidence that a first correlation between the
mechanical loss angle and the doping of the Ion Beam Sputtered tantala coatings has been observed, and future work in
this area aims to use this information to develop coatings with improved loss. There is also significant research into
alternative high index of refraction materials including Hafnia, Zirconia, Niobia and silicon nitride
\cite{Whitepaper:2012}. Furthermore, the use of doping  to further lower the mechanical loss, and stabilise the coatings
under the application of higher annealing temperatures ($>600^{\circ}\textrm{C}$) may also lead to additional
improvements. Any reduction of the high index material will also require a parallel approach to studying lower
mechanical loss low index layers, which although are lower, cannot be neglected. The study of slow annealing processes
which reduce the total stress in the coating layers appears to be a possible area of future R\&D, in addition to doping
in order to stabilise the coatings against crystallisation. Conservative assumptions suggest that a further factor of 2
in the loss angle may be achieved, giving a total gain of $\simeq3.2$ when combined with larger beam diameter.

\subsubsection{Overall improvement from warm upgrades}
\begin{figure}
\begin{center}
\includegraphics[width=1.0\textwidth]{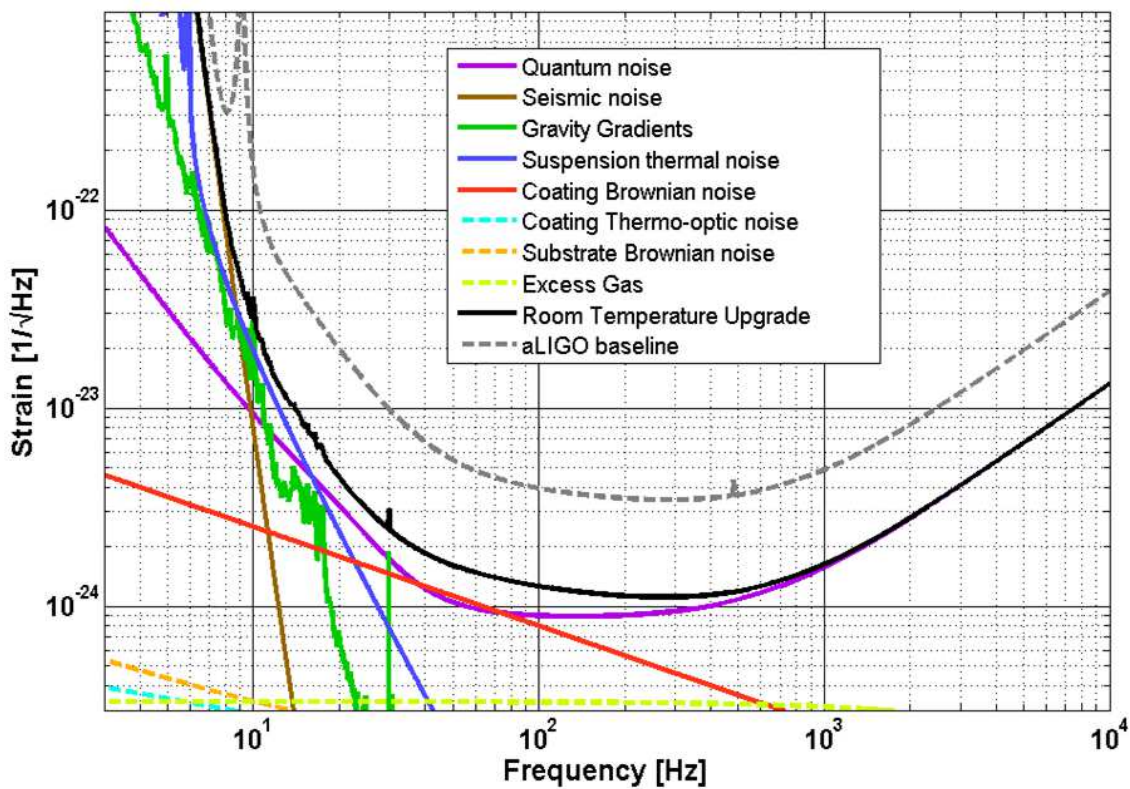}
\caption{The strain sensitivity improvement for a room temperature upgrade based on fused silica \cite{Red:2012}}.
\label{red_warm}
\end{center}
\end{figure}

With the improvements stated above a warm upgrade could potentially offer a factor of 3 improvement in the strain
sensitivity as shown in Figure~\ref{red_warm}. This would correspond to an improved event rate by a factor of 27
compared to the advanced detector network. Preliminary analysis \cite{Red:2012} has shown that such
an upgrade can 
be obtained for a fraction of the hardware cost of a new infrastructure, such as the Einstein
Telescope.

\subsection{Cryogenic upgrades}

Cryogenic upgrades assume a combined approach of lowering the mechanical loss angle and the temperature to provide
improved thermal noise performance \cite{Punturo:2010, Adhikari:2013, Hammond:2012}. While amorphous materials (e.g.\
fused silica) work extremely well at room temperature, a low temperature dissipation peak at around 40\,K limits their
use for cryogenic applications. Crystalline materials on the other hand, such as silicon and sapphire, do not show such
broad dissipation peaks and are the best candidates \cite{Nowick:1972, Punturo:2010}. An additional benefit of
crystalline materials is that they display high thermal conductivity in the region
$>500\,\textrm{W}\textrm{m}^{-1}\textrm{K}$ which is important for both reducing the problems associated with thermal
lensing with high laser powers and extracting heat from the optic. Silicon is the baseline material for the European
Einstein Telescope design study. A significant benefit is the fact that large sample sizes will become available within
the next 10 years (e.g. $>50$\,cm diameter for 200\,kg optics) as a result of the semiconductor industry. However,
further R\&D is required to develop 1550\,nm laser technology to allow transmissive interferometry. Sapphire is the
material which is the baseline for the KAGRA detector currently being built in Japan. While the maximum diameter of the
available sample, and thus the ultimate mass of the optic, is lower than silicon, a benefit is that sapphire can be
operated at 1064 nm with transmissive optics.

One choice which is important to make is the operating temperature of the detector. Silicon exhibits a zero in its thermal expansion coefficient at both 120\,K and 18\,K \cite{Punturo:2010} and thus the thermoelastic noise contribution can be made negligible at this point (with remaining loss terms due to surface and bulk). Operating at 120\,K allows significant  heat ($\simeq 10$ W) to be extracted via radiative cooling to a nearby
cold structure \cite{Adhikari:2013} although the thermal noise performance is not necessarily as good as a low
temperature operation (assuming identical laser power). On the other hand, operating at 18 K offers high gain in terms
of thermal noise performance, but with the challenge of needing to extract several tens of milliwatts of heat along the
suspension fibres. The deposited heat originates from the absorption of the incident laser beam onto the mirror coatings
(ideally a few ppm absorption) and also optical absorption of the laser beam as it travels through the transmissive
optic of the interferometer (a few ppm/cm).

\subsubsection{Suspension thermal noise}
Significant modelling has been undertaken on the silicon suspension thermal noise for both 120\,K and 18\,K operation
\cite{Cumming:2013, Hammond:2012}. Similar to warm upgrades, long suspension elements with thick attachment ends
to optimise the stored energy into the thin fibre are necessary. There has already been work to test fabrication
techniques which typically fall into two categories. The first requires the use of a heated pedestal growth technique
\cite{Alshourbagy:2006} to pull a crystal fibre out of a melt. Challenges involved with this techniques include
maintaining temperature stabilisation of the molten silicon bath and ensuring that the resulting crystal is single axis
and not polycrystalline. The second method assumes etching or machining the fibres out of wafers which results in
rectangular geometry suspension elements. Challenges with this method include making strong Silicon fibres which retain
their strength after etching  or mechanical machining.

\begin{figure}
\begin{center}
\includegraphics[width=1.0\textwidth]{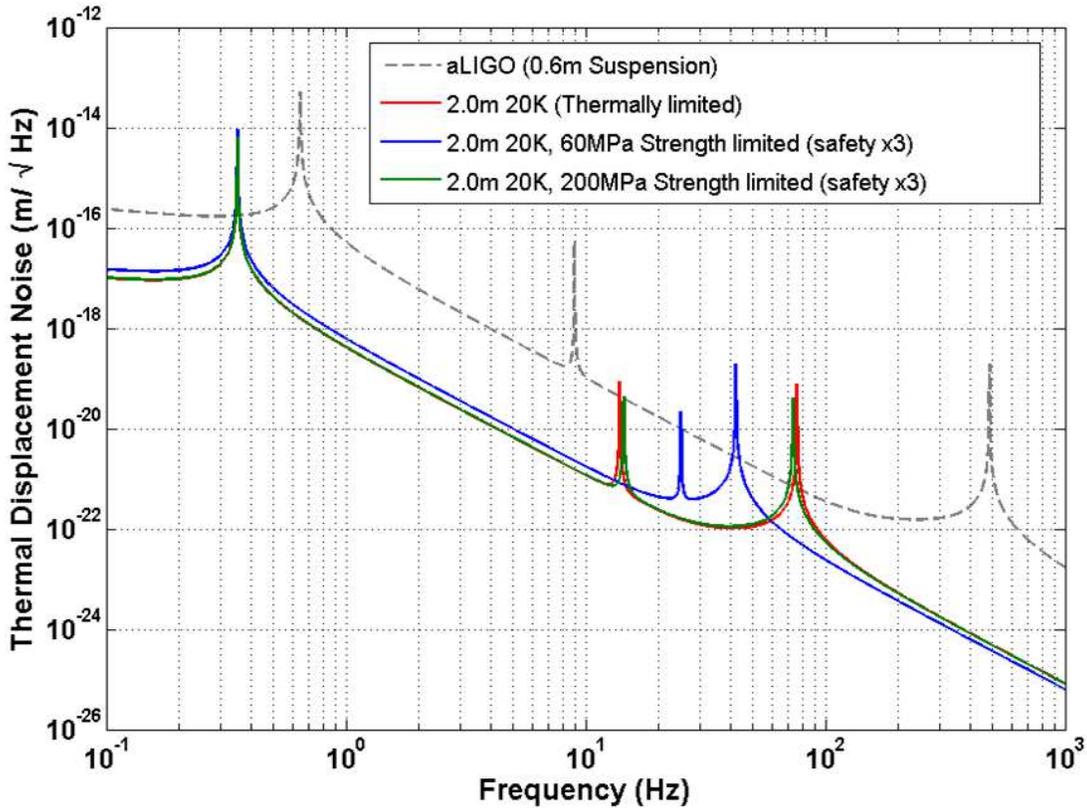}
\caption{The suspension thermal noise performance for a variety of silicon suspensions which are both strength and
thermally limited.}
\label{silicon}
\end{center}
\end{figure}

As stated above, fabricating crystalline suspensions requires a careful analysis of the strength and thermal
requirements of the support elements, in addition to reproducible bonding techniques \cite{Dari:2010, Veggel:2009}.
Current measurements \cite{Cumming:2013} of the tensile strength of silicon elements yield values in the region
200-500\,MPa, although further work to improve this value is very likely. From these measurements it is possible to show
that a 200\,kg test mass suspension would be currently limited by strength requirements. If silicon could be made
sufficiently strong, then the suspension elements would be reduced in cross sectional area, ultimately improving the
suspension thermal noise via improved dissipation dilution. It is important to note that for low temperature operation
(e.g. $<50$\,K), where radiative cooling is not viable, there is a further limitation that the elements must remain of
sufficient cross-section to remove the deposited heat. Figure~\ref{silicon} shows an example of the thermal noise
performance of a silicon suspension comprising a 200\,kg test mass and 2\,m long suspension elements. For reference the
aLIGO design is also shown. Three silicon suspensions are shown: a thermally limited suspension (red line), a strength
limited suspension operating at 60\,MPa with a safety factor of 3 (blue line) and a strength limited suspension
operating at 180\,MPa with a safety factor of 3 (green line). The peaks correspond to the pendulum mode, the vertical
mode and only the first violin mode (other modes at harmonics have been omitted). There is significant improvement
potential for such suspensions operating at cryogenic temperature, with up to a factor of 50 improvement in the thermal
noise performance at 10\,Hz. Operating at higher stress values pushes apart the vertical mode and violin mode which is
desirable. However, assuming a maximum stress of 200\,MPa (60\,MPa with a safety factor of 3) shows that strength of the
silicon is currently the driving parameter.

\subsubsection{Coating thermal noise }
There are significant gains to be made from coating thermal noise improvements. Again the use of heavier test masses
allows the possibility of increasing the beam size by a factor of 1.6. As noted above amorphous coatings are not ideal
at cryogenic temperatures due to the observed low temperature loss peak. Thus alternative coatings are currently under
development based on crystalline materials which are grown with Molecular Beam Epitaxy. Two such coatings technologies
are based on gallium arsenide (GaAs) and gallium phosphide (AlGaP). The AlGaAs coatings, whose high/low index layers 
are achived by varying the concentration of aluminium (Al$_{0.92}$Ga$_{0.08}$As/GaAs for low/high index respectively),
have  been studied in detail in opto-mechanical experiments \cite{Cole:2013}. These coatings offer  losses which are
at the level of $4\ee{-6}$ and absorption less than 6\,ppm. A challenge is that the coatings are grown on a gallium
arsenide substrate and thus have to be lifted off onto either silicon or sapphire substrates. Such techniques are
reasonably standard in the microfabrication industry and have already been used to put coatings onto crystalline
substrates which have been temperature cycled to 10\,K and lower. There is further possibility to scale these up to
larger substrates, although additional R\&D to put these coatings onto curved mirror substrates and test their
mechanical loss on representative size samples is necessary. The GaP coatings, whose high/low index layers are also
achieved by varying the concentration of aluminium (Al$_{0.9}$Ga$_{0.1}$P/GaP for low/high index respectively), are
lattice matched to silicon and can be grown directly onto these substrates with minimum defects. An upper limit on the 
losses of these coatings is $<2\ee{-4}$ although work is ongoing to further measure this relatively new coating
technology. The optical absorption is still too high for current interferometers, although the possibility of annealing
these coatings may lead to a possible reduction. There is also significant research ongoing to study resonant waveguide
mirrors \cite{Rosenblatt:1997, Bruckner:2010, PhysRevD.88.042001}. These devices work by resonantly coupling incident
light into a periodically structured grating and have the benefit of removing all coating layers. Recent loss
measurements suggest that at cryogenic temperatures the loss can be an order of magnitude lower than that of the Ion
Beam Sputtered coatings and absorption is at the level of $\simeq 100\,\textrm{ppm}$. Again a combined approach of
alternative coating technologies and materials is likely to deliver gains of up to a factor of 5--10 in mechanical loss.
The coating thermal noise varies with the square root of the mechanical loss, as detailed in equation \ref{CTN}, and
when combined with the increased beam diameter, a significant gain of 8--16 in coating thermal noise can be expected.

\subsection{Dual detectors}

\begin{figure}
\begin{center}
\includegraphics[width=1.0\textwidth]{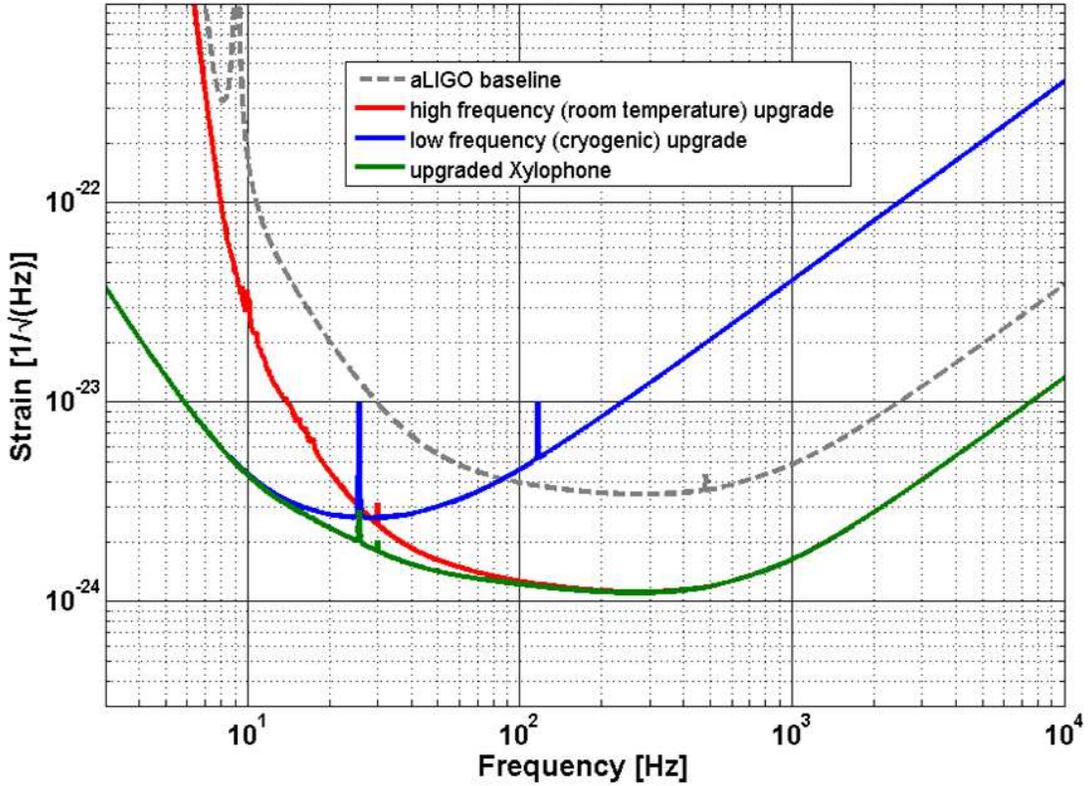}
\caption{The strain sensitivity improvement for a cryogenic upgrade based on silicon at 20 K. Note that for the blue and
green trace only thermal  noise and quantum noise are considered. To achieve this level of sensitivity methods have to
be implemented to eliminate Newtonian noise and seismic noise correspondingly.}
\label{red_cold}
\end{center}
\end{figure}

There are challenges associated with operating a cryogenic detector at high laser power. Thus an interesting concept
includes the use of a dual operating scheme or Xylophone concept \cite{Shoemaker_xylo, Conforto2004228, Hild2010a}. This
would comprise a high frequency, room temperature detector, operating at high power and utilising fused silica as the
baseline material. In parallel, a cryogenic detector, operating with low power and utilising crystalline silicon or
sapphire, would form the low frequency detector. Such a configuration could provide the benefits of both operating
schemes with broadband improvement. Figure~\ref{red_cold} shows the strain sensitivity improvement for a Xylophone
configuration. Only the fundamental noise sources are shown on this plot (quantum noise, suspension thermal noise,
coating thermal noise) and thus such a detector would require a significant improvement or subtraction of the Newtonian
noise background and seismic noise. When compared to aLIGO, the improvement in the strain sensitivity is $\simeq 7$ at
frequencies less than $20$\,Hz.

\subsection{Astrophysics}
It is very likely that the first direct detections of \gws will happen with the operation of aLIGO and AdV. However,
following this vital milestone there are many questions in astrophysics, fundamental physics and cosmology that
can be addressed through further observations \cite{Sathyaprakash:2009} in particular with the huge numbers of sources
expected for ET \cite{Sathyaprakash:2012}. At their design sensitivities aLIGO and AdV should have astrophysical reaches
of a few hundred Mpc for coalescing compact binary sources. As mentioned in Section~\ref{sec:gwsearches} it is estimated
that they should expect to observe $\sim 40$ coalescences of binary neutron stars per year \cite{Abadie:2010b} when they
reach their design sensitivity. They could also potentially observe tens of binary stellar mass
black hole systems and neutron
star-black hole binaries, which are systems that have never been observed through electromagnetic observations. Whilst
with ET thousands of these sources could be observed out to high cosmological redshifts \cite{Sathyaprakash:2012}. Through
their use as ``standard sirens'' \cite{Schutz:1986} their distances can be determined and, provided a measure of the
redshift can be estimated (see below), they can be used to determine cosmological parameters.

This also opens up the prospect of multimessenger astronomy in which \gw and electromagnetic, or high energy
neutrino/cosmic-ray, observations can be combined to get the most information about a source (see \cite{Marka:2011} for
a review of multimessenger efforts with initial detectors, and \cite{Chassande-Mottin:2011} for how it could be used in
the ET era). As \gw detectors are sensitive to the whole sky the localisation of short transient sources requires
triangulation using signal arrival times in multiple detectors. With the two aLIGO instruments and AdV it is expected
that typically a source may be localised to tens of square degrees on the sky \cite{Fairhurst:2011, Aasi:2013}. This
localisation would be further improved though the inclusion of KAGRA and/or the siting of one aLIGO instrument in India
\cite{Indigo, Unnikrishnan:2013}. Localising a \gw source could allow it to be unambiguously associated with an
electromagnetic transient, such as a $\gamma$-ray burst or supernova, providing unique insight into the mechanisms for
these astrophysical events. Localising inspiral events could identify their host galaxies, allowing the redshift to be
measured and enabling their use as cosmological probes.

There are many further exciting prospects that could arise from these observations, including probing neutron star
equations of state and testing GR in extreme gravity situations. Many of these are described in more detail in
e.g.\ \cite{Sathyaprakash:2009, Sathyaprakash:2012}. This is an extremely exciting time for the field of \gw
astrophysics with the advanced detectors coming online in 2015, first observations expected $\geq 2016$ and a strong
international programme of R\&D for future upgrades and detectors.

\section*{Acknowledgements}
The authors gratefully acknowledge the support of the United States National Science Foundation for the construction and
operation of the LIGO Laboratory and the Science and Technology Facilities Council of the United Kingdom, the
Max-Planck-Society, and the State of Niedersachsen/Germany for support of the construction and operation of the GEO600
detector. The authors also gratefully acknowledge the support of the research by these agencies and by the Australian
Research Council, the Council of Scientific and Industrial Research of India, the Istituto Nazionale di Fisica Nucleare
of Italy, the Spanish Ministerio de Econom\'ia y Competitividad, the Conselleria d'Economia Hisenda i Innovaci\'o of the
Govern de les Illes Balears, the Royal Society, the Scottish Funding Council, the Scottish Universities Physics
Alliance, The National Aeronautics and Space Administration, the Carnegie Trust, the Leverhulme Trust, the David and
Lucile Packard Foundation, the Research Corporation, and the Alfred P. Sloan Foundation. S.\ H.\ acknowledges the support
from the European Research Council (ERC-2012-StG: 307245). M.\ P.\ is funded by the Science \& Technology Facilities Council grant no.\ ST/L000946/1. This paper has been assigned LIGO document no.\
LIGO-P1300081. 
\bibliographystyle{tMOP}
\bibliography{jmo_article}

\end{document}